\documentclass[12pt]{article}
\usepackage{graphicx}
\usepackage{epstopdf}
\usepackage{epsfig}
\usepackage{amssymb}
\usepackage{setspace}
\usepackage{caption}
\usepackage{amsfonts}
\usepackage{color}
\usepackage{amsmath}
\usepackage{float}
\usepackage{comment}
\newcommand {\be}{\begin{equation}}
\newcommand {\ee}{\end{equation}}
\newcommand {\bea}{\begin{array}}
	
	\newcommand {\eea}{\end{array}}

\newcommand{\RN}{Reissner-Nordstrom\,}
\newcommand{\SCH}{Schwarzschild\,}

\textwidth=6.5in \hoffset=-.75in \voffset=-1in \numberwithin{equation}{section}
\numberwithin{figure}{section}

\begin{document}

\begin{titlepage}
\vspace{1cm} 
\begin{center}
{\Large \bf {On Destroying Extremal Magnetized Black Holes}}\\
\end{center}
\vspace{2cm}
\begin{center}
\renewcommand{\thefootnote}{\fnsymbol{footnote}}
Haryanto M. Siahaan{\footnote{haryanto.siahaan@gmail.com; haryanto.siahaan@unpar.ac.id}}
\\
Center for Theoretical Physics,\\ Physics Department, Parahyangan Catholic University,\\
Jalan Ciumbuleuit 94, Bandung 40141, Indonesia
\renewcommand{\thefootnote}{\arabic{footnote}}
\end{center}

\begin{abstract}
The gedanken experiment by Wald to destroy a black hole using a test particle in the equatorial plane is adopted to the case of extremal magnetized black holes. We find the presence of external magnetic fields resulting from the ``Ernst magnetization'' permits a test particle to have strong enough energy to destroy the black hole. However, the corresponding effective potentials show that such particles would never reach the horizon.
\end{abstract}
\end{titlepage}\onecolumn 
\bigskip 

 \section{Introduction}
\label{sec:intro}

According to general relativity, nothing can stop a massive enough star to collapse into a singularity after its nuclear fuel runs out. However, according to the cosmic censorship conjecture introduced by Penrose, the singularity resulting from a total gravitational collapse must be hidden behind a surface known as the black hole horizon\cite{Wald:1984rg.book}. Since the birth of this conjecture, a large number of studies have been devoted to test its validity. For example the one by Wald who proposed a gedanken experiment to overspin or overcharge a Kerr-Newman black hole by throwing a test particle with some specific physical properties from far infinity into the black hole. By capturing the particle, it is expected that the black hole would pass its extremal bound and becomes a naked singularity.

An infalling test particle with mass $m$, charge $q$, and angular momentum $L$ into a black hole could transform the black hole into a naked singularity in Kerr-Newman spacetime, provided that the particle's energy $E$ satisfies
\be \label{EboundUp}
\left( {M + E} \right)^2  < \left( {Q + q} \right)^2  + \left( {\frac{{\left| J \right| + \left| L \right|}}{{M + E}}} \right)^2\,.
\ee
The last expression is simply the violation of extremal bound for a Kerr-Newman black hole. In the inequality above, $M,~Q$ and $J$ are the mass, charge, and
angular momentum a Kerr-Newman black hole. In addition to the maximum energy constraint (\ref{EboundUp}), the test particle must also experience an attractive effective potential in its all way down to the black hole horizon. This is guaranteed by the condition $V_{eff} < 0 ~\forall~ r > r_+$, or equivalently reads ${\dot r}^2 > 0~ \forall ~r > r_+$, where $r_+$ is the outer event horizon of the black hole \cite{Wald:1984rg.book}. Nevertheless, there exists a minimum energy required by the test particle to reach the horizon $r_+$ \cite{Misner:1974qy}
\be \label{EboundDn}
E_{\min }  = \left. {\frac{{q\left( {g_{t\phi } A_\phi   - g_{\phi \phi } A_t } \right) - g_{t\phi } L}}{{g_{\phi \phi } }}} \right|_{r = r_ +  } < E\,,
\ee 
dictated by the corresponding geodesic equation for the test particle. Note that, in an asymptotically flat solution in Einstein-Maxwell theory, $E$ and $L$ above are the conserved quantities associated to the Killing vectors $\partial_t$ and $\partial_\phi$. Interestingly, the magnetized spacetime discussed in this paper also possess these Killing vectors, thus one would expect to find the quantities similar to $E$ and $L$ above. There are several attempts reported in literature to define these conserved quantities in magnetized spacetimes \cite{Gibbons:2013dna,Astorino:2015naa,Booth:2015nwa,Astorino:2016hls}. In particular, we employ the method presented in \cite{Astorino:2016hls} to define the mass, charge, and angular momentum of magnetized black holes, as reviewed in App. \ref{app.Mass}. A test particle may have an energy which obeys both (\ref{EboundUp}) and (\ref{EboundDn}), and such particle may break the horizon of black holes.

However, in the Wald's thought experiment, Kerr-Newman family of black holes in extremal states cannot be overspun or overcharged to pass their extremality. This is due to the fact that the maximum and minimum energies required by the test particle coincide each other. Interestingly, it was pointed out later on that if the Wald's gendanken experiment is performed starting from the near-extremal condition, then such black holes could be destroyed by a test particle. In the case of near-extremal \RN black hole, it was Hubeny \cite{Hubeny:1998ga} who first showed that the black hole could jump the extremality by capturing a charged test particle thrown from far away. Subsequently, a decade later Jacobson and Sotiriou reported that a quite similar conclusion can be drawn in the case of a near-extremal Kerr black hole\cite{Jacobson:2009kt}, where the captured neutral test particle brings angular momentum. Obviously, it is straightforward to ask whether the near-extremal charged and rotating black holes can be overspun and or overcharged to pass their extremality adopting the scenario by Hubeny \cite{Hubeny:1998ga} or Jacobson-Sotiriou \cite{Jacobson:2009kt}. This was addressed in several works, for example refs. \cite{SaaPRD} and \cite{Siahaan:2015ljs} to the case of Kerr-Newman and Kerr-Sen black holes respectively. 

Nonetheless, the studies mentioned above on the possibility of turning the near-extremal black holes into naked singularities neglect the self-force, self-energy, and radiative effects. If one considers these aspects, for example as in the works \cite{CardosoPRL,CardosoPRD,Zimmerman:2012zu}, the Wald thought experiment cannot be the mechanism to transform a black hole into a naked singularity. Moreover, several studies even take into account the quantum effects in discussing the cosmic censorship. For example, quite recently the authors of \cite{Semiz} reported that quantum effects support the cosmic censorship conjecture, while in \cite{Casadio:2015sda} the authors discussed a quantum version of this conjecture. We mention this work to show that cosmic censorship conjecture is still debatable and of great interest to researchers. In case of eventually nature allows the existence of a naked singularity, even some physics related to that have been explored \cite{Virbhadra}. 

In the last couple of years, there is a growing interest in the studies of black holes interacting with external magnetic fields. One of the reasons is that the astronomers found some evidences of strong magnetic fields in the center of galaxies, where normally supermassive black hole sits. Considering that the external magnetic fields are just some perturbations in the spacetime, Wald introduced a gravito-electromagnetic system where the metric is still Kerr and the corresponding vector fields are constructed using a definite linear combination of the Killing vectors \cite{Wald:1974np}. As the magnetic fields get stronger, clearly the spacetime will be affected, and the solution by Wald cannot be employed any further. In strong magnetic fields, an exact solution by Ernst and Wild \cite{ernst-wild} can be used to model the magnetized black hole. They employed a Harrison-like transformation \cite{Harrison} to get a magnetized solution from a known unmagnetized one, which sometime is called as the seed solution, in the Einstein-Maxwell theory. Setting the mass, electric charge, and rotation of the black holes in the Ernst-Wild solution \cite{ernst-wild} to be vanished, one get the Melvin universe \cite{Melvin:1963qx} which is a static non-singular cylindrically symmetric spacetime that contains an axial homogeneous magnetic field aligned in the $z$-axis. 

The solution by Ernst and Wild \cite{ernst-wild} which describes magnetized black boles could be more popular if only it is asymptotically flat. Moreover, if this is the case then one can simply use the standard textbook prescriptions to define the physical quantities in the magnetized spacetime, for example the ADM formulation of mass and energy \cite{Wald:1984rg.book}. Despite this lacking of asymptotic flatness, people find that the solution by Ernst and Wild \cite{ernst-wild} is adequate to model the high energetic supermassive black hole in the center of galaxies. This black hole is surrounded by some external magnetic fields as a resulting from the rotating matters around it \cite{Aliev:1989wx}. Therefore, it is not surprising that research reports on aspects of the magnetized black holes are continuously appeared in literature \cite{Brito:2014nja,Bicak:2015lxa,Shoom:2015uba,Kolos:2015iva,Tursunov:2014loa,Shiose:2014bqa,Azreg-Ainou:2016tkt,Tursunov:2016dss,Astorino:2015lca,Astorino:2016hls,Siahaan:2015xia,Orekhov:2016bpc,Booth:2015nwa,Gibbons:2013yq,Gibbons:2013dna,Rogatko:2016knj}. Particularly, the mass, electric charge, angular momentum, and entropy of this black hole are studied thoroughly in \cite{Gibbons:2013dna,Booth:2015nwa,Astorino:2016hls}. 

Assuming that the Enrst-Wild solution, or the properly modified version, describing black holes immersed by some magnetic fields is good enough to model the astrophysical black holes influenced by strong external magnetic fields, one may wonder how the presence of external magnetic fields may contribute to the possibility of destroying such black hole by adopting the Wald gedanken experiment \cite{waldAP1974}. This is the question that we would like to answer in this paper. The method is straightforward, by using the techniques in \cite{Gao:2012ca} and \cite{Siahaan:2015ljs}. However, the mathematical works are quite involved due to the complexity of Ernst-Wild solution. Therefore, at some points we will depend on the numerical results in drawing the conclusions.

The organization of this paper is as follows. In section \ref{s.1} we provide some reviews on magnetization transformation to obtain a magnetized spacetime. Subsequently, in section \ref{sec.MB} we provide some necessary details on the magnetized black holes studied in this paper. Then in section \ref{s.2} we study some properties of a test particle moving in the equatorial plane outside the magnetized black hole backgrounds. In this section we obtain the maximum and minimum energies needed by the test particle to break the black hole's horizon, and also the leading terms of the corresponding effective potentials to tell whether the capturing process may take place or not. In section \ref{s.3} we provide some numerical examples to support the results presented in the section \ref{s.2}. Finally, the discussions and conclusions are given in section \ref{s.discussion}. The unit system that we use in this paper is $c = G = \hbar = 1$. 


\section{Magnetization of black holes}\label{s.1}

In 1976, Ernst and Wild \cite{ernst-wild} reported\footnote{Following the method found by Ernst previously that year \cite{Ernst}, hence sometime will be referred as the Ernst magnetization.} an exact solution describing a Kerr-Newman black hole immersed by a homogeneous magnetic field. They employed a Harrison-like transformation \cite{Harrison} to the Kerr-Newman fields,
\be \label{Lewis1}
ds^2  = f\left( {d\phi  - \omega dt} \right)^2  - {{f^{-1}\Delta _r y^2}  dt^2  + e^{2\mu } \left( {\frac{{dr^2 }}{{\Delta _r }} + \frac{{dx^2 }}{{y^2 }}} \right)}\,,
\ee
and
\be \label{vector1}
{\bf A} = A_t dt + A_\phi  d\phi \,.
\ee
The resulting magnetized version reads\footnote{Note that the changes are only in $f$ and $\omega$, where $g_{rr}$ and $g_{xx}$ are left unchanged.}
\be \label{Lewis2}
d\tilde s^2  = \tilde f\left( {d\phi  - \tilde \omega dt} \right)^2  - {{{\tilde f}^{-1}\Delta _r y^2}  dt^2  + e^{2\mu } \left( {\frac{{dr^2 }}{{\Delta _r }} + \frac{{dx^2 }}{{y^2 }}} \right)}\,,
\ee
and
\be \label{vector2}
{\bf \tilde A} = \tilde A_t dt + \tilde A_\phi  d\phi \,.
\ee
In the equations above, the corresponding coordinate is $x^\mu = \left(t,r,x,\phi\right)$ and for the sake of simplicity we have used $y^2 = 1-x^2$. The functions $f$, $\omega$, $\Delta_r$, $\Delta_x$, $\mu$, $A_t$, and $A_\phi$, together with the corresponding ``tilde'' versions depend on the coordinates $r$ and $x$ only. Hence, it is obvious that the line elements (\ref{Lewis1}) and (\ref{Lewis2}) are expressed in the Lewis-Papapetrou form, which is appropriate to the spacetime with $\partial_t$ and $\partial_\phi$ Killing vectors. The Kerr-Newman metric (\ref{Lewis1}) and the magnetized one (\ref{Lewis2}) enjoy these symmetries, and both can be studied using the Ernst formalism \cite{Ernst:1967wx,Ernst:1967by}.

Explicitly, the functions appearing in the line element (\ref{Lewis1}) describing Kerr-Newman spacetime are
\be\label{f_comp_KN}
f = \frac{{\left( {\left( {r^2  + a^2 } \right)^2  - \Delta _r a^2 y^2} \right)y^2}}{{r^2  + a^2 x^2 }}\,,
\ee
\be \label{e_2mu_KN}
e^{2\mu }  = r^2  + a^2 x^2 
\ee
\be \label{omega_KN}
\omega  = \frac{{\left( {r^2  + a^2  - \Delta _r } \right)a}}{{\left( {r^2  + a^2 } \right)^2  - \Delta _r a^2 y^2}}\,,
\ee 
\be\label{Delr_KN}
\Delta _r  = r^2  + a^2  + Q^2  - 2Mr\,,
\ee
where the corresponding vector field is
\be \label{A_KN}
{\bf A} = \frac{{Qr\left( {dt - ay^2 d\phi } \right)}}{{r^2  + a^2 x^2 }}\,.
\ee 
In general relativity, the Kerr-Newman solution describes the spacetime outside an object with mass $M$, charge $Q$, and angular momentum $J=Ma$.


The Ernst magnetization relates the functions $\tilde{f}$ and $\tilde{\omega}$ contained in the metric (\ref{Lewis2}) to $f$ and $\omega$ in (\ref{Lewis1}). In the followings, we will review the relations. In his seminal papers, Ernst showed that the Einstein-Maxwell equations for a stationary and axial symmetric spacetime can be equivalently written as the Ernst equations\footnote{See \cite{IslamBook} for a quite extensive review on the subject.} \cite{Ernst:1967wx,Ernst:1967by},
\be \label{ErnstEq1}
f\nabla ^2 {\cal E} = \left( {\nabla {\cal E} - 2\Phi^* \nabla \Phi} \right)\cdot \nabla {\cal E}\,,
\ee 
\be \label{ErsntEq2}
f\nabla ^2 {\Phi} = \left( {\nabla {\cal E} - 2\Phi^* \nabla \Phi} \right)\cdot \nabla {\Phi}\,.
\ee 
Explicitly, the operator $\nabla$ reads
\be\label{nabla} 
\nabla  = \frac{{\sqrt {\Delta _r {y^2} } \left\{ {\left( {2r - M} \right)\partial _r  - x\partial _x } \right\} + i\left\{ {{y^2} \Delta _r '\partial _x  + 2x\Delta _r \partial _r } \right\}}}{{\left( {r - M} \right){y^2} \Delta _r ' + 2x^2 \Delta _r }}\,,
\ee
where $\Delta_r '$ is the derivative of $\Delta_r$ with respect to $r$. In the equations above, $\cal E$ is the Ernst gravitational potential defined with respect\footnote{Since the line element is shown in general as $
ds^2  = f\left( {d\phi  - \omega dt} \right) + hdt^2  + g_{ij} dx^i dx^j$. It is understood that $f,\omega,h$ and $g_{ij}$ are functions of $x_1$ and $x_2$, and the indices $i,j =1,2$.} to the Killing vector $\partial_\phi$ \cite{Stephani:2003tm}, i.e.
\be \label{ErntsPotE}
{\cal E} = f + \left| \Phi  \right|^2 +i\varphi\,,
\ee 
where $\varphi$ is the twist potential. The corresponding Ernst electromagnetic potential reads 
\be \label{ErntsPotPh}
\Phi = A_\phi + i B_\phi\,,
\ee 
and $B_\phi$ is related to the Maxwell fields (\ref{vector1}) as 
\be \label{nablaA_tToApBp}
\nabla A_t  =  - \omega A_\phi   - i\sqrt{\Delta _r{y^2}} f^{ - 1} B_\phi\,.
\ee 

The magnetized gravitational and electromagnetic Ernst potentials associated to the metric (\ref{Lewis2}) and the vector (\ref{vector2}) satisfy the same form of Ernst equations (\ref{ErnstEq1}) and (\ref{ErsntEq2}),
\be \label{ErnstEq1T}
{\tilde f}\nabla ^2 \tilde{\cal E} = \left( {\nabla \tilde{\cal E} - 2\tilde\Phi^* \nabla \tilde\Phi} \right)\cdot \nabla \tilde{\cal E}\,,
\ee 
\be \label{ErsntEq2T}
{\tilde f}\nabla ^2 \tilde{\Phi} = \left( {\nabla \tilde{\cal E} - 2\tilde\Phi^* \nabla \tilde\Phi} \right)\cdot \nabla \tilde{\Phi}\,.
\ee
The Harisson-like transformation to magnetize the ``old'' Ernst potentials is given by
\be \label{HarissonForErnstPots}
\tilde{\cal E} = \Lambda ^{ - 1} {\cal E} ~~,~~
\tilde{\Phi}  = \left( {\Phi  - \frac{{B{\cal E}}}{2}} \right)\Lambda ^{ - 1} \,,
\ee
where
\be \label{Lambda}
\Lambda  = 1 - B\Phi  + \frac{{B^2 {\cal E}}}{4}\,.
\ee
The parameter $B$ is a constant which represents the strength of external magnetic fields involved. Accordingly, the functions in (\ref{Lewis2}) are related to those in (\ref{Lewis1}) through the followings
\be\label{tildef}
\tilde f = f\left| \Lambda  \right|^{ - 2} \,,
\ee 
\be \label{tildeomega}
\nabla \tilde \omega  = \left| \Lambda  \right|^2 \nabla \omega  - \frac{\sqrt{\Delta _r{y^2}} }{f}\left( {\Lambda ^* \nabla \Lambda  - \Lambda \nabla \Lambda ^* } \right)\,,
\ee
and the other functions such as $\mu$ and $\Delta_r$ remain unchanged. The fact that $\Delta_r$ does not change due to the transformation (\ref{HarissonForErnstPots}) indicates that the horizons of black holes are similar, at least in the equatorial plane\footnote{It is shown that the presence of external magnetic fields in Ernst-Wild spacetime deform the horizon to have an egg-shape appearance \cite{Booth:2015nwa}.}, to that of unmagnetized case. Moreover, it also tells us that the bound of extremality for black holes under discussions after the magnetization procedure remains unaffected.

As a matter of fact, Wald \cite{Wald:1974np} had performed some analysis of Kerr black holes in the background of uniform external magnetic field before the work of Ernst on magnetized black holes \cite{Ernst,ernst-wild}. In Wald's work, the corresponding vector potential $A^\mu$ is generated from a superposition of the Killing vectors $\xi^\mu_{\left(t\right)}$ and $\xi^\mu_{\left(\phi\right)}$, from which the external test magnetic field can be obtained. However, the spacetime metric in Wald's analysis is still the ordinary Kerr, meaning the external test magnetic field has no influence to the spacetime\footnote{We notice that the work of Shaymatov et. al. \cite{Shaymatov:2014dla}, where the authors addressed an issue closely related to the one presented in our paper, uses the vector potential given in \cite{Wald:1974np} and the spacetime under consideration is unmagnetized Kerr.}. Interestingly, in the work of Ernst \cite{Ernst}, which then was extended by Ernst and Wild \cite{ernst-wild}, the external magnetic field is not just a test field, it deforms the spacetime. It is obvious from the equations (\ref{tildef}) and (\ref{tildeomega}), we can notice that the external magnetic fields affect the spacetime through $\Lambda$, that appears explicitly in the line element (\ref{Lewis2}). Taking the limit $B \to 0$ in (\ref{Lewis2}), i.e. $\Lambda\to 1$, the magnetized spacetime (\ref{Lewis2}) reduces to the unmagnetized one (\ref{Lewis1}) as expected. 

We note that, as it was also frequently mentioned in the papers which discuss the magnetized black holes in the framework of Ernst, working with magnetized spacetimes in Einstein-Maxwell family is quite demanding task, and in particular the magnetized Kerr-Newman. This is because the analytic expressions for the corresponding fields, which must satisfy the source free Einstein-Maxwell equations, are quite elaborate. Hence, it would be a tedious work to perform the verification of some available solutions in the literature without using any Symbolic Manipulation Programs, such as MAPLE or MATHEMATICA. From our surveys in literature, we find the exact solutions reported by Aliev and Galtsov \cite{Aliev:1989wz} to be reliable\footnote{In particular, this work is part of the series work works by the authors in magnetized black holes \cite{Aliev:1980et,Aliev:1986wu,Aliev:1988wy,Aliev:1989sw,Aliev:1989wx,Aliev:1989wz,Galtsov:1978ag}.}. Therefore, in the following two subsections we will make use of their results to study some aspects of the magnetized versions of the \RN and Kerr black holes.

\section{Magnetized black holes}\label{sec.MB}

In the previous section, we have reviewed briefly the magnetization prescription according to Ernst \cite{Ernst} in Einstein-Maxwell theory. This section is devoted to present some aspects of magnetized \RN and Kerr solutions. Several improvements of the corresponding metric and vector fields, for example scaling the $\phi$ coordinate to avoid a conical singularity \cite{Aliev:1989wz} and introducing a constant part in the $\phi$ component of vector field to yield $A_\phi \left(x = \pm 1 \right) = 0$ as in \cite{Astorino:2015naa}, are considered. In this section, and also for the rest of this paper, we make distinctions between the mass, angular momentum, and electric charge of the magnetized black holes as reviewed in App. \ref{app.Mass} to the mass, angular momentum, and electric charge parameters brought from the ``seed'' solutions. Mass, angular momentum, and electric charge of a magnetized black hole are denoted by ``tilde'', i.e. $\tilde M$, $\tilde J$, and $\tilde Q$ respectively, while the parameters are without ``tilde''.

\subsection{Magnetized Reissner-Nordstrom black holes}

The ordinary \RN spacetime reduces to the famous \SCH solution if the electric charge is turned off. This black hole has two horizons, which coincide in the extremality. Due to the black hole electric charge, the interaction between a test particle and black hole is not gravitational only, but also electromagnetic if the particle is electrically charge. Now, one can imagine that the spacetime is also filled by some external magnetic fields, thence the interaction between a charged test particle with the external magnetic fields will add to the total interactions. Some new features should come up, and the discussions are worth to do since the astrophysical black holes observed in the sky might be surrounded by some external magnetic fields produced by moving charges around the black holes. Some aspects of the magnetized \RN black hole will be reviewed in the followings.

We start by writing the generic \RN metric, obtained by setting $a=0$ in (\ref{Lewis1}), i.e.
\be \label{metricRN}
ds^2  =  - r^{ - 2} \Delta _{RN} dt^2  + r^2 \left( {\Delta _{RN}^{ - 1} dr^2  + y^{ - 2} dx^2  + y^2 d\phi ^2 } \right)
\ee
where $\Delta _{RN} = r^2+Q^2-2Mr$, and the corresponding vector field is
\be \label{vecRN}
{\bf A} = r^{-1}Q dt\,.
\ee 
Since later we will have the notions of mass, electric charge, and angular momentum for a magnetized \RN black hole, which are distinguishable from the mass, electric charge, and angular momentum of Kerr-Newman black holes, then $M$ and $Q$ appeared in (\ref{metricRN}) would be called as the mass and charge parameters, respectively. The mass, angular momentum, and charge of a magnetized \RN black hole will depend on these parameters.

The Ernst potentials associated to the \RN solution (\ref{metricRN}) and (\ref{vecRN}), dictated by the equations (\ref{ErntsPotE}) and (\ref{ErntsPotPh}), can be read as
\be \label{ErnstPotRN}
{\cal E} = r^2 y^2 + Q^2 x^2 ~~~,~~~ \Phi  = -iQx\,.
\ee
The magnetization transformation (\ref{HarissonForErnstPots}) yields a pair of Ernst potentials which correspond to the magnetized \RN spacetime, which can be written as
\be\label{EpsilonMRN} 
\tilde{\cal E} = \left( {r^2 y^2 + Q^2 x^2 } \right)\Lambda_{RN}^{ - 1} \,,
\ee
and
\be \label{PhiMRN}
\tilde\Phi  = -\left( {iQx +\frac{B}{2}\left( {r^2 y^2 + Q^2 x^2 } \right)} \right)\Lambda_{RN}^{ - 1} \,.
\ee
In the two potentials above, 
\be
\Lambda_{RN}  = 1 + \frac{{B^2 }}{4}\left( {r^2 y^2 + Q^2 x^2 } \right) + iBQx\,.
\ee 
Accordingly, the corresponding line element (\ref{Lewis2}) that corresponds to the Ernst potential (\ref{EpsilonMRN}) contains
\be 
\tilde \omega = \tilde \omega _{RN}  =  \frac{{2BQ}}{r} - B^3 Qr - \frac{{B^3 Q^3 }}{{2r}} + \frac{{B^3 Q\left( {r^2  - 2Mr + Q^2 } \right)y^2}}{{2r}}\,,
\ee
\be 
\tilde f =\tilde f_{RN} = r^2 y^2\left| {\Lambda _{RN} } \right|^2 ~~,~~\Delta _r  = \Delta _{RN}~~,~~ e^{2\mu }  = r^2 \,.
\ee
Note that $\tilde \omega _{RN}$ vanishes in the absence of external magnetic field, i.e. the metric becomes \RN, as one would expect. Moreover, it is interesting to point out that there appears a new feature when one discusses the magnetized \RN spacetime, which does not exist in the ordinary \RN case, namely the dragging effect denoted by $\tilde \omega _{RN}$. Accordingly, one can interpret that an infalling test particle would experience a dragging effect as it gets closer the black hole due to the presence of external magnetic fields. 

Nevertheless, Hiscock\footnote{He showed that the periodicity of angular coordinate $\phi$ is not $2\pi$ anymore.} \cite{Hiscock:1980zf} reported that the original magnetized solution by Ernst and Wild \cite{Ernst,ernst-wild} suffers a conical singularity problem. One way to cure this problem is by performing the ``scaling'' \cite{Aliev:1989wx} $\phi\to \phi ' =  \left| {\Lambda _{RN,0} } \right|^2 \phi$, where 
\be \label{Lambda0RN}
\left| {\Lambda _{RN,0} } \right|^2 = \left| {\Lambda _{RN} \left(x=1\right) } \right|^2 = {1 + \frac{{3B^2 Q^2 }}{2} + \left( {\frac{{BQ}}{2}} \right)^4 }\,.
\ee
As a result of this scaling, the modified magnetized \RN metric now reads 
\be \label{Lewis2scaledRN}
d\tilde s^2  = \tilde f_{RN}\left( {\left| {\Lambda _{RN,0} } \right|^2 d\phi  - {\tilde \omega}_{RN} dt} \right)^2  - {\Delta _{RN} {{\tilde f_{RN}}^{-1} y^2}} dt^2  + e^{2\mu } \left( {\frac{{dr^2 }}{{\Delta _{RN} }} + \frac{{dx^2 }}{{{y^2} }}} \right)\,.
\ee

Obviously, the vector field solution in the magnetized \RN system is not the same as that in the case of generic \RN solution anymore. The associated vector field in magnetized case contains not only the timelike component, but also a spacelike one which reads
\be \label{AtMRN}
A_\phi   = \left| {\Lambda _{RN,0} } \right|^2  \left({\mathop{\rm Re}\nolimits} \tilde \Phi  + A_{\phi 0} \right)\,,
\ee 
where $\tilde \Phi$ is given in (\ref{PhiMRN}). It resembles the vector fields in Kerr-Newman or Kerr-Sen solutions \cite{Siahaan:2015ljs}, where ${\bf A} = A_t dt +A_\phi d\phi$. This resemblance can be understood from the fact that both Kerr-Newman and Kerr-Sen are electrically charged and rotating, thence the metric contains a non-vanishing $g_{t\phi}$ which happens also to be the case of magnetized \RN solution. Particularly, the constant 
\be 
A_{\phi 0}  =  \frac{{2BQ^2 \left( {12 + B^2 Q^2 } \right)}}{{B^4 Q^4  + 24B^2 Q^2  + 16}}
\ee
have been added in (\ref{AtMRN}) to guarantee that $A_\phi\left(x=\pm 1\right)$ vanishes \cite{Astorino:2015lca}. Furthermore, explicitly the timelike component of $\bf A$ reads
\be \label{ApMRN}
A_t  = -\frac{Q}{r} + \frac{3}{2}B^2 Qr + \frac{3}{4}\frac{{B^2 Q^3 }}{r} + \frac{{3y^2\Delta _{RN} }}{{4r}}B^2 Q - {\tilde \omega}_{RN} {\mathop{\rm Re}\nolimits} \tilde \Phi \,.
\ee

Now we have already established the metric and vector field solutions describing the magnetized \RN spacetime. For the future practical purpose, the components of metric tensor describing the magnetized \RN spacetime $d\tilde s^2  = \tilde g_{\mu \nu } dx^\mu  dx^\nu$ can be expressed explicitly as
\be\label{gttMRN}
{\tilde g}_{tt}  =  - \frac{{\Delta _{RN} \left| {\Lambda _{RN} } \right|^2  - r^4 {\tilde \omega}_{RN} ^2 y^2}}{{r^2 \left| {\Lambda _{RN} } \right|^2 }}\,,
\ee
\be
{\tilde g}_{t\phi }  =  - \frac{{r^2 {\tilde \omega}_{RN} y^2\left| {\Lambda _{RN,0} } \right|^2 }}{{\left| {\Lambda _{RN} } \right|^2 }}\,,
\ee
\be
{\tilde g}_{rr}  = \frac{{r^2 \left| {\Lambda _{RN} } \right|^2 }}{{\Delta _{RN} }}\,,
\ee
\be
{\tilde g}_{xx}  = \frac{{r^2 \left| {\Lambda _{RN} } \right|^2 }}{{y^2 }}\,,
\ee
\be\label{gppMRN}
{\tilde g}_{\phi \phi }  = \frac{{r^2 \Delta _x \left| {\Lambda _{RN,0} } \right|^4 }}{{\left| {\Lambda _{RN} } \right|^2 }}\,.
\ee
The metric above together with the vector components (\ref{ApMRN}) and (\ref{AtMRN}) will be employed to calculate the geodesic of a charged test particle moving towards a magnetized \RN black hole in section \ref{s.2} and the corresponding effective action.

Black holes can exist in this magnetized \RN spacetime, whose horizons take the same radii to that of the unmagnetized case, i.e. $r_ \pm   = M \pm \sqrt {M^2  - Q^2 } $. Accordingly, the extremal condition is also the same, namely when $Q=M$. However, since the magnetized \RN spacetime is not asymptotically flat, defining the conserved quantities such as the mass, electric charge, and angular momentum in this spacetime cannot be done by using the ADM formalism. In app. \ref{app.Mass}, we highlight the results presented in \cite{Astorino:2016hls} which we are used to define conserved quantities in magnetized spacetime discussed in this paper. By setting the angular momentum parameter to be vanished in (\ref{massMKN}), (\ref{JMKN}), and (\ref{QMKN}), one gets
\be\label{massMRN}
\tilde M_{RN}  = \left[ {M^2  + \left( {\frac{3}{2}M^2  - Q^2 } \right)Q^2 B^2  + \frac{{M^2 Q^4 B^4 }}{{16}}} \right]^{\frac{1}{2}} \,,
\ee
\be\label{JMRN}
\tilde J_{RN}  =  - \left( {1 + \frac{{Q^2 B^2 }}{4}} \right)BQ^3 \,,
\ee
and
\be\label{QMRN}
\tilde Q_{RN}  = \left( {1 - \frac{{Q^2 B^2 }}{4}} \right)Q\,,
\ee
as the mass, angular momentum, and charge of a magnetized \RN black hole. Note that the presence of external magnetic fields gives rise to a rotation in the magnetized \RN spacetime, denoted by the angular momentum (\ref{JMKN}), and this effect vanishes when the external magnetic field is turned off.

\subsection{Magnetized Kerr}

Now let us briefly review some aspects of magnetized Kerr spacetime. We start from the unmagnetized Kerr whose line element can be written as
\[
ds^2  = \frac{{\left( {\left( {r^2  + a^2 } \right)^2  - \Delta _K a^2 y^2 } \right)y^2 }}{{r^2  + a^2 x^2 }}\left( {d\phi  - \frac{{2Mradt}}{{\left( {\left( {r^2  + a^2 } \right)^2  - \Delta _K a^2 y^2 } \right)}}} \right)^2 
\]
\be 
- \frac{\left({r^2  + a^2 x^2 }\right) \Delta _K y^2 dt^2}{{\left( {\left( {r^2  + a^2 } \right)^2  - \Delta _K a^2 y^2 } \right)y^2 }} +\frac{\left(r^2+a^2 x^2\right)dr^2}{\Delta_K} +\frac{\left(r^2+a^2 x^2\right)dx^2}{y^2} \,,
\ee 
where $\Delta_K = r^2+2Mr-a^2$. Accordingly, the corresponding Ernst gravitational potential related to the Kerr metric above is
\be\label{ErnstKerr} 
{\cal E}_{K} = \left( {r^2  + a^2 } \right)y^2 - 2iaM\left( {3 - x^2 } \right)x + \frac{{2a^2 y^4 M}}{{r + iax}}\,,
\ee
while the associated Ernst electromagnetic potential $\Phi$ is zero.

Following the prescription (\ref{HarissonForErnstPots}) by Ernst, the function
\be \label{LambdaKerr}
\Lambda _{K}  = 1 + \frac{{B^2 {\cal E}_{K} }}{4}\,,
\ee 
will be used to transform (\ref{ErnstKerr}) to a magnetized version of Kerr spacetime. The Ernst potentials as a result of this magnetization process takes the form
\be \label{EK}
\tilde {\cal E}_{K}  = \Lambda _{K}^{ - 1} {\cal E}_{K} = \frac{{4\left( {r^2  + a^2 } \right)^2 y^2  - 8iaM\left( {3 - x^2 } \right)x + \frac{{8a^2 y^4 M}}{{r + iax}}}}{{4 + B^2 \left( {\left( {r^2  + a^2 } \right)^2 y^2  - 2iaM\left( {3 - x^2 } \right)x + \frac{{2a^2 y^4 M}}{{r + iax}}} \right)}}\,,
\ee 
and
\be \label{PhiK}
\tilde \Phi _{K}  =  - \frac{{B{\cal E}_{K} }}{{2\Lambda _{K} }} =  - \frac{{B\left( {\left( {\left( {r^2  + a^2 } \right)^2 y^2  - 2iaM\left( {3 - x^2 } \right)x + \frac{{2a^2 y^4 M}}{{r + iax}}} \right)} \right)}}{{2 + \frac{1}{2}\left( {\left( {\left( {r^2  + a^2 } \right)^2 y^2  - 2iaM\left( {3 - x^2 } \right)x + \frac{{2a^2 y^4 M}}{{r + iax}}} \right)} \right)}}\,.
\ee 
Explicitly, the metric tensor components that belong to the magnetized Kerr spacetime $d\tilde s^2  = \tilde g_{\mu \nu } dx^\mu  dx^\nu$ can be expressed as
\be\label{gttKerr}
{\tilde g}_{tt}  =  - \frac{{\left( {r^2  + a^2 x^2 } \right)\left| \Lambda_K  \right|^2 \Delta _{K} }}{{\left( {r^2  + a^2 } \right)^2  - a^2 y^2\Delta _{K} }} + \frac{{\left( {\left( {r^2  + a^2 } \right)^2  - \Delta _{K} a^2 y^2} \right)y^2\tilde \omega _{K}^2 }}{{\left( {r^2  + a^2 x^2 } \right)\left| \Lambda_K  \right|^2 }}
\ee
\be
{\tilde g}_{rr}  = \frac{{\left( {r^2  + a^2 x^2 } \right)\left| \Lambda_K  \right|^2 }}{{\Delta _{K} }}
\ee
\be
{\tilde g}_{xx}  = \frac{{\left( {r^2  + a^2 x^2 } \right)\left| \Lambda_K  \right|^2 }}{{y^2}}
\ee
\be
{\tilde g}_{t\phi }  =  - \frac{{\left( {\left( {r^2  + a^2 } \right)^2  - \Delta _{K} a^2 y^2} \right)y^2\left| {\Lambda _{K,0} } \right|^2 \tilde \omega _{K} }}{{\left( {r^2  + a^2 x^2 } \right)\left| \Lambda_K  \right|^2 }}
\ee
and
\be\label{gppKerr}
{\tilde g}_{\phi \phi }  = \frac{{\left( {\left( {r^2  + a^2 } \right)^2  - \Delta _{K} a^2 y^2} \right)y^2 \left| {\Lambda _{K,0} } \right|^4 }}{{\left( {r^2  + a^2 x^2 } \right)\left| \Lambda_K  \right|^2 }}
\ee
where 
\be\label{Lambda0K} \left| {\Lambda _{K,0} } \right|^2 = \left| {\Lambda _{K} } \left(x=1\right) \right|^2 = 1+ B^4 M^2 a^2 \,.
\ee 
This factor $\left| {\Lambda _{K,0} } \right|^2$ which appears in the magnetized Kerr spacetime,
\be \label{Lewis2scaledKerr}
d\tilde s^2  = \tilde f\left( {\left| {\Lambda _{K,0} } \right|^2 d\phi  - {\tilde \omega}_{K} dt} \right)^2  - {\Delta _{K} {{\tilde f}^{-1} y^2}} dt^2  + e^{2\mu } \left( {\frac{{dr^2 }}{{\Delta _{K} }} + \frac{{dx^2 }}{{{y^2} }}} \right)\,,
\ee
removes the conical singularity problem in the original magnetized Kerr metric by Ernst \cite{Ernst}. The vector field components which accompany the magnetized Kerr metric above can be written as\footnote{We have considered the factor $\left| {\Lambda _{K,0} } \right|^2$ in $A_\phi$ following the scaling $\phi \to \phi ' = \left| {\Lambda _{K,0} } \right|^2\phi$ in the metric.}
\[A_\phi   = \left| {\Lambda _{K,0} } \right|^2 \left(A_{\phi 0}  + {\mathop{\rm Re}\nolimits} \left\{ {\tilde \Phi _{K} } \right\}\right) \]
\be \label{ApKerr}
= \left| {\Lambda _{K,0} } \right|^2 \left(A_{\phi 0}  - \frac{{B\left( {{\mathop{\rm Re}\nolimits} \left\{ {\tilde {\cal E}_{K} } \right\}{\mathop{\rm Re}\nolimits} \left\{ {\Lambda _{K} } \right\} + {\mathop{\rm Im}\nolimits} \left\{ {\tilde {\cal E}_{K} } \right\}{\mathop{\rm Im}\nolimits} \left\{ {\Lambda _{K} } \right\}} \right)}}{{2\left| \Lambda_K  \right|^2 }}\right)\,,
\ee
and
\[A_t  = \left( {\frac{{\Delta _{K} y^4 \left( {r^3  - 3a^2 r + 2Ma^2 } \right) - 8r}}{4} - \frac{{\Delta _{K} y^2\left( {4a^2 M^2 r - r\left( {r^2  + a^2 } \right)} \right)}}{{\left( {r^2  + a^2 } \right)^2  - \Delta _{K} a^2 y^2}}} \right)aMB^3\]
\be \label{AtKerr}
 - \tilde \omega _{K} {\mathop{\rm Re}\nolimits} \left\{ {\tilde \Phi _{K} }\right\}\,.
\ee

Likewise, the constant
\be\label{Ap0-Kerr}
A_{\phi 0}  = \frac{{2B^3 a^2 M^2 }}{{1 + B^4 M^2 a^2 }}
\ee
in (\ref{ApKerr}) guarantees that $A_\phi \left(x = \pm 1\right)$ vanishes \cite{Astorino:2015naa}. The associated ${\tilde \omega}_K$ to get the explicit expression for the function $A_t$ above is given by \cite{ernst-wild}
\be \label{wKerr}
\tilde \omega _K  = \frac{{\left( {\alpha  - \beta \Delta _K } \right)}}{{r^2  + a^2 }}\,,
\ee
where
\be 
\alpha  = a\left( {1 - B^4 a^2 M^2 } \right)\,,
\ee 
and
\[
\beta  = \frac{{a\left( {r^2  + a^2 x^2 } \right)}}{{\left( {r^2  + a^2 } \right)^2  - \Delta _K a^2 y^2 }} + \left( {\frac{B}{2}} \right)^4 \left( {\frac{{4M^2 a^3 x^2 \left\{ {\left( {r^2  + a^2 } \right)\left( {3 - x^2 } \right)^2  - 4a^2 y^2 } \right\}}}{{\left( {r^2  + a^2 } \right)^2  - \Delta _K a^2 y^2 }}} \right.
\]
\be 
\left. { + 2Ma\left( {\frac{{a^2 y^6 \left\{ {\left( {r^2  + a^2 } \right)r + 2Ma^2 } \right\}}}{{\left( {r^2  + a^2 } \right)^2  - \Delta _K a^2 y^2 }} - 3ry^4  - 4rx^2 \left( {3 - x^2 } \right)} \right)} \right)\,.
\ee
Another metric function $\tilde{f}_K$ can be obtained using the relation ${\mathop{\rm Re}\nolimits} \left\{ {{\tilde{\cal E}}_K} \right\} = \tilde{f}_K + \left| {\tilde \Phi}_K \right|$.

Obviously, the mathematical expression for the full magnetized Kerr spacetime is significantly more involved compared to the unmagnetized version, or even to the familiar Kerr-Newman solution. Moreover, this magnetized Kerr spacetime can also contain black holes, where the black hole horizons radii take the same value as that in the unmagnetized one, i.e. $r_ \pm   = M \pm \sqrt {M^2  - a^2 } $. Consequently, the extremal condition will be also the same to that in a generic Kerr solution, namely $a=M$. Related to the mass, angular momentum, and charge of massive body in this magnetized Kerr spacetime, again one can obtain these quantities by setting the charge parameter $Q$ to be vanished in (\ref{massMKN}), (\ref{JMKN}), and (\ref{QMKN}), i.e.
\be\label{massMK}
\tilde M_{K}  = \left[ {M^2  + 2 J^2 B^2  + J^2 M^2 B^4} \right]^{\frac{1}{2}} \,,
\ee
\be\label{JMK}
\tilde J_{K}  =  \left( {1 - J^2 B^4} \right) J \,,
\ee
and
\be\label{QMK}
\tilde Q_{K}  = 2JB \,.
\ee
Note that the presence of external magnetic fields yields a massive body in this magnetized Kerr spacetime to have charge given by eq. (\ref{QMK}) which vanishes at $B=0$. 

\section{Test particles and the naked singularity}\label{s.2}

\subsection{Energy and angular momentum}
The motion of a classical test particle in a general curved background follows the geodesic equation \cite{waldAP1974}
\be\label{eq.motion.incurved}
\frac{{du^\mu  }}{{ds}} + \Gamma _{\alpha \beta }^\mu  u^\alpha  u^\beta   = \frac{q}{m}F^{\mu \nu } u_\nu \,,
\ee 
where $m$ and $q$ are the mass and electric charge parameters of the particle respectively. It is well known that the geodesic equation (\ref{eq.motion.incurved}) can be obtained alternatively from the Lagrangian \cite{Misner:1974qy,waldAP1974}
\be\label{eq.Lang}
{\cal L} = \frac{1}{2}mg_{\alpha \beta } {\dot x}^\alpha {\dot x}^\beta + qA_\mu  {\dot x}^\mu\,.
\ee 
Accordingly, since the spacetime is stationary and axisymmetric, there are two constants of motion related to the Lagrangian (\ref{eq.Lang}), namely the energy $E$ and angular momentum $L$. These quantities are given by
\be\label{E}
E =  - \frac{{\partial {\cal L}}}{\partial {\dot t}} =  - m\left( {g_{tt} \dot t + g_{t\phi } \dot \phi } \right) - qA_t\, ,
\ee 
and
\be \label{L}
L = \frac{{\partial {\cal L}}}{\partial {\dot \phi}} = m\left( {g_{t\phi } \dot t + g_{\phi \phi } \dot \phi } \right) + qA_\phi \,,
\ee 
respectively. Note that the quantities $m$, $q$, $E$, and $L$ above are the conserved ones that belong to the test particle which is a probe in the magnetized spacetime under consideration. Therefore, it is not necessary to assign some conserved charges $\tilde m$, $\tilde q$, and $\tilde L$, associated to the test particle following the formulas in App. \ref{app.Mass} since the correction terms due to the presence of $B$ parameter would be negligibly small. In other words, $\tilde m \simeq m$, $\tilde q \simeq q$, and $\tilde L \simeq L$, where $m$, $q$, and $L$ are those that appear in (\ref{E}) and (\ref{L}).

In general, a magnetized spacetime metric takes the form
\be 
ds^2  = g_{tt} \left( {r,\theta } \right) dt^2  + g_{rr} \left( {r,\theta } \right) dr^2  + g_{\theta \theta } \left( {r,\theta } \right) d\theta ^2  + g_{\phi \phi } \left( {r,\theta } \right) d\phi ^2  + 2g_{t\phi } \left( {r,\theta } \right) dtd\phi
\ee
which is similar to the general form of Kerr-Newman line element, thence one can obtain the relation between $E$ and $L$ from (\ref{E}) and (\ref{L}) as
\be \label{EtoL}
E = \frac{{g_{t\phi } }}{{g_{\phi \phi } }}\left( {qA_\phi   - L} \right) - qA_t  + {\sqrt {{\left( {\frac{{g_{t\phi }^2  - g_{\phi \phi } g_{tt} }}{{g_{\phi\phi }^2 }}} \right)\left( {\left( {L - qA_\phi  } \right)^2  + m^2 g_{\phi \phi } \left( {1 + g_{rr} \dot r ^2 + g_{\theta \theta } \dot \theta ^2} \right)} \right)}}}\,.
\ee
In obtaining the last equation we have imposed the timelike condition ${\dot x}_\mu{\dot x}^\mu = -1$. Moreover, in getting the equation (\ref{EtoL}) we consider the solution that implies ${\dot t} > 0$ only. In the next two subsections, we will make use of the general expression of energy above in exploring the possibility of turning an extremal magnetized black hole into a naked singularity, by letting the black hole captures a test particle equipped with some specific initial physical conditions. 

In the followings, we will discuss the maximum and minimum bounds of energy for the test particle to break an extremal magnetized black hole. It is already known that an extremal Kerr-Newman, or the static and neutral limits of it, cannot be broken in the Wald's gedanken experiment where the test particle moving in the equatorial plane. This is because a non-zero positive $\Delta E \equiv E_{\max}-E_{\min}$ so a test particle could break the black hole does not exist if the initial condition of the black hole is extremal. The story is different if the initial state of black holes is near-extremal rather than extremal \cite{Hubeny:1998ga,Jacobson:2009kt}, where an infalling test particle may push the black hole to pass its extremality. Interestingly, we will find out that a test particle under the influence of some external magnetic fields resulting from the Ernst magnetization can have a positive $\Delta E$, even though the black holes is extremal. 


\subsection{Test particles in the magnetized \RN spacetime}\label{ss.TRN}

We start by evaluating equation (\ref{EtoL}) in the extremal condition $Q=M$ for magnetized \RN black holes evaluated at $r=r_+$. The energy $E$ obtained is then interpreted as the minimum one required by a test particle to reach the horizon of an extremal magnetized \RN black hole. Interestingly, in such set up the squared root term in (\ref{EtoL}) vanishes and the final expression is considerably simplified,
\be\label{EminMRN}
E_{\min}  = \frac{{q\left\{ {64 + 3B^2 M^2 \left( {B^4 M^4  - 28B^2 M^2  - 80} \right)} \right\}}-32BL\left(3B^2 M^2 -4\right)}{{64 + 4B^2 M^2 \left( {24 + B^2 M^2 } \right)}}\,.
\ee
Now let us obtain the maximum energy of the test particle which leads to the destruction of black hole's horizon. As it is shown in \cite{Astorino:2016hls}, the conserved quantities in magnetized Kerr-Newman spacetime are $\tilde M$, $\tilde J$, and $\tilde Q$, which stand for the mass, angular momentum, and charge of an object in magnetized spacetime respectively. Furthermore, these quantities are functions of $B$ parameter, and the parameters of black holes in the corresponding seed solution, namely $M$, $Q$, and $J$. Moreover, a magnetized \RN black hole achieves an extremal state when\footnote{See app. \ref{app.Mass} for a discussion on how this relation can be obtained.} $\tilde M^4  = \tilde Q^2 \tilde M^2  + \tilde J^2$, or equivalently $M=Q$ as indicated in (\ref{extremeMKN}) by setting $J=0$. Accordingly, the maximum energy of the test particle which falls into the back hole to allow the production of naked singularity in magnetized \RN spacetime can be read from (\ref{viol.ext}), namely
\be\label{EmaxMRN}
E_{\max }  = \frac{1}{2}\sqrt {2\left( {\tilde Q_{RN} } + q \right)^2  + 2\sqrt {\left( {\tilde Q_{RN} } +q \right)^4  + 4\left( {\tilde J_{RN} } + L \right)^2 } }  - \tilde M_{RN} \,.
\ee
Employing equations (\ref{massMRN}), (\ref{QMRN}), and (\ref{JMRN}) together with extremal condition $Q=M$ to (\ref{EmaxMRN}) yield
\[
E_{\max }  = \frac{{\sqrt 2 }}{2}\left( {\sqrt {\left( {M + q - \frac{{M^3 B^2 }}{4}} \right)^4  + 4\left( {BM^3  + \frac{{M^5 B^3 }}{4} + L} \right)^2 } } \right.
\]
\be 
\left. { + \left( {M + q - \frac{{M^3 B^2 }}{4}} \right)^2 } \right)^{\frac{1}{2}}  - M - \frac{{M^3 B^2 }}{4}\,.
\ee

If there exists a gap between $E_{\min}$ in (\ref{EminMRN}) and $E_{\max}$ in (\ref{EmaxMRN}), i.e.  then one may interpret that the captured test particle can lead to the production of a naked singularity from a MRN black hole. Evaluating $\Delta E \equiv E_{\max} - E_{\min}$ analytically to show that it can be non-zero positive is quite complicated task to do. Nonetheless, since we can Taylor expand $\Delta E$ in $B$ parameter and consider up to the second order only. This is possible due to the small numerical value of $B$ if it is compared to $M$. Furthermore, for the sake of simplicity, let us consider the energy
\be\label{EmaxMRNxcal}
{\cal E}_{\max }  = \frac{1}{2}\sqrt {2\left( {\tilde Q_{RN} } \right)^2  + 2\sqrt {\left( {\tilde Q_{RN} } \right)^4  + 4\left( {\tilde J_{RN} } \right)^2 } }  - \tilde M_{RN} < E_{\max}\,,
\ee
instead of $E_{\max}$ in (\ref{EmaxMRN}), and Taylor expand $\Delta {\cal E} \equiv {\cal E}_{\max} - E_{\min}$ in $B$ up to the second order,
\be\label{DelEappMRN}
\Delta {\cal E} \approx  - q - 2LB + \frac{{21}}{4}qM^2 B^2 \,.
\ee 
This $\Delta {\cal E}$ above can be positive for a range of $B$, which will be verified in the numerical plot below. From this result, one can say that there is a chance for the test particle under some circumstances to break the horizon of extremal magnetized \RN black hole.

\begin{figure}
	\begin{center}
		\includegraphics[scale=0.4]{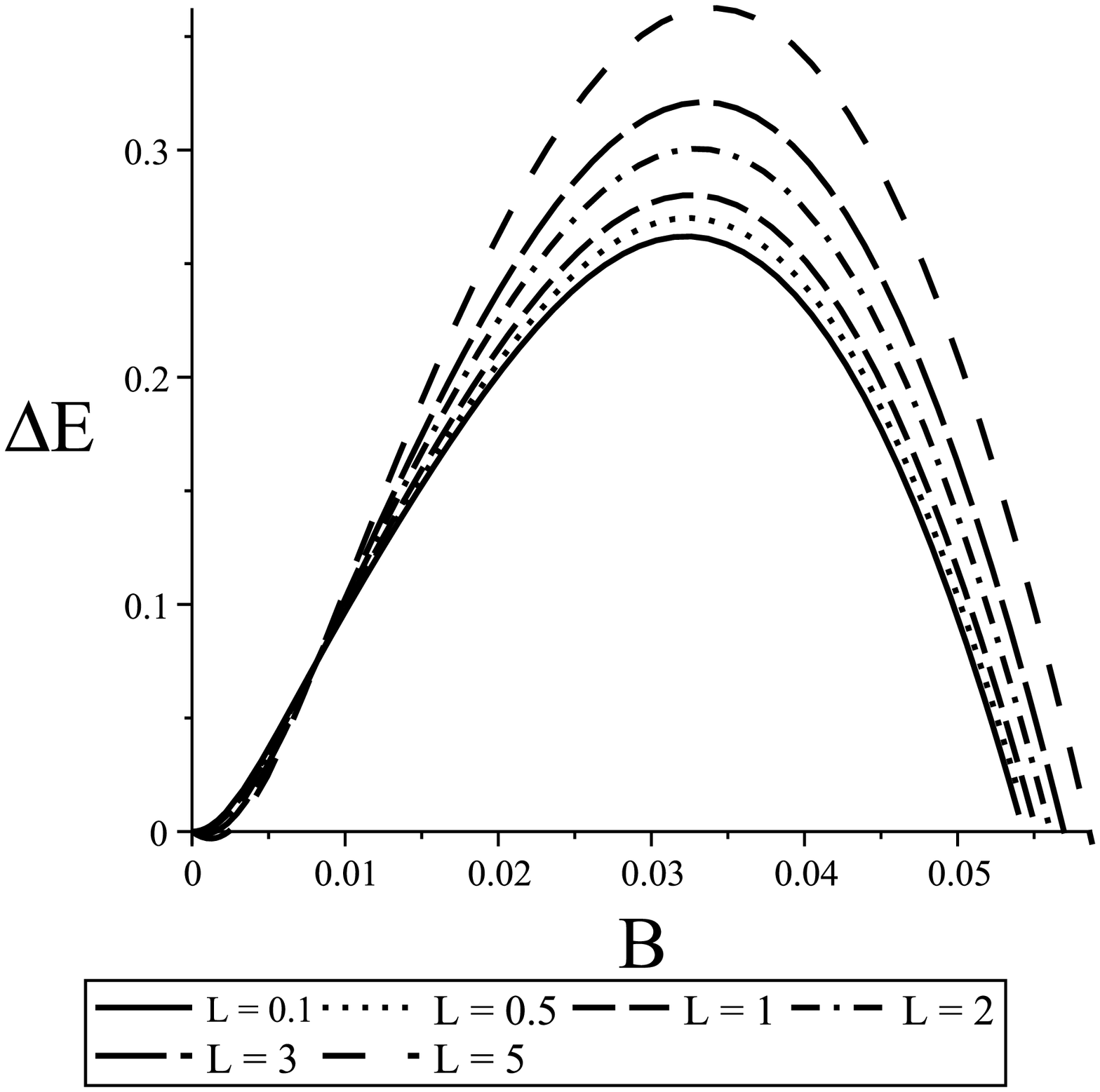}
	\end{center}
	\caption{Plots of $\Delta E = E_{\max} -E_{\min}$ for $M=100$, $q=0.05$, and several numerical values of charge parameter $L$.} \label{DelE6plotsLvar}
\end{figure}

\begin{figure}
	\begin{center}
		\includegraphics[scale=0.4]{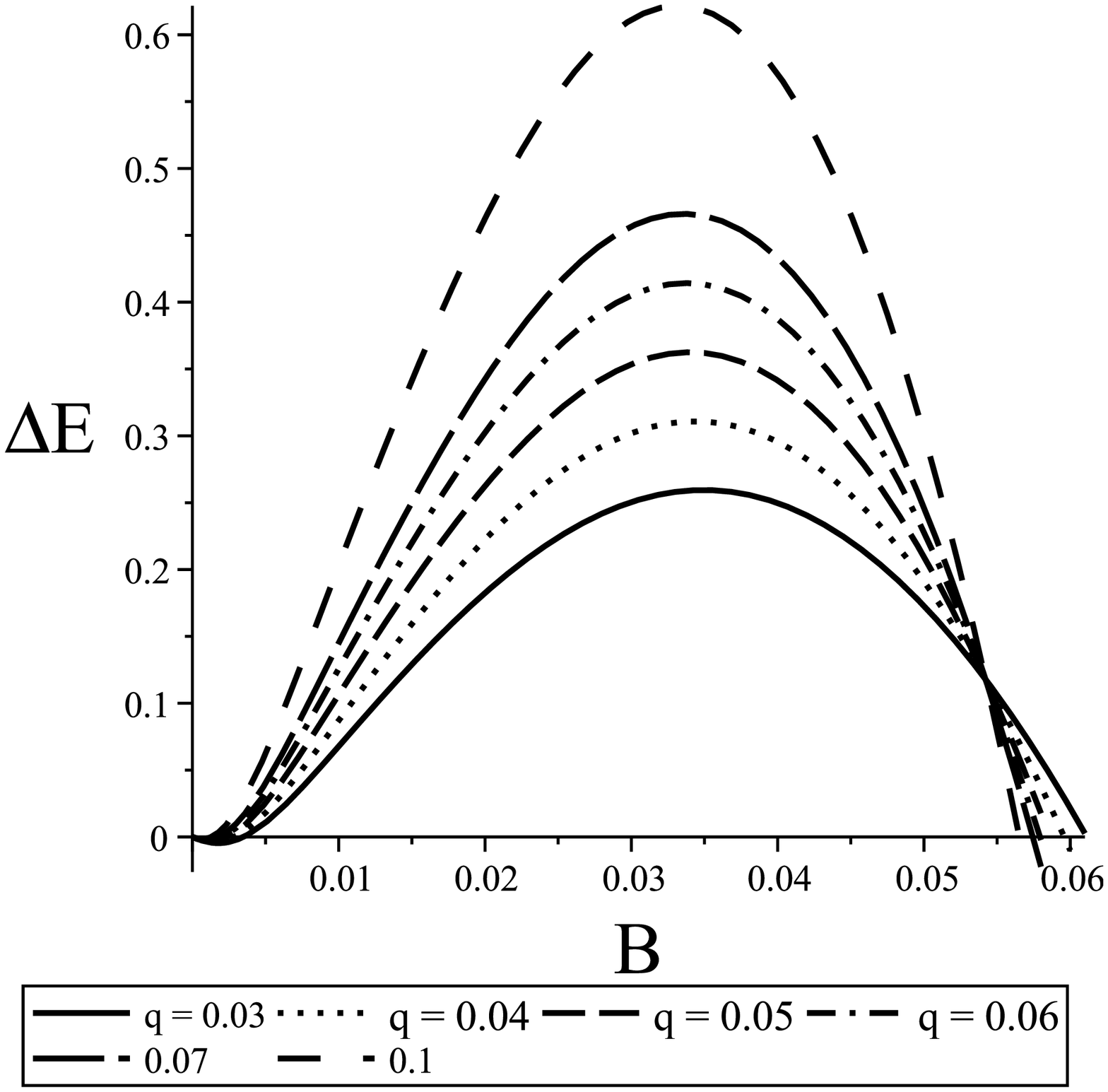}
	\end{center}
	\caption{Plots of $\Delta E = E_{\max} -E_{\min}$ for $M=100$, $L=5$, and several numerical values of charge parameter $q$.} \label{DelE6plotsQvar}
\end{figure}

\begin{figure}
	\begin{center}
		\includegraphics[scale=0.4]{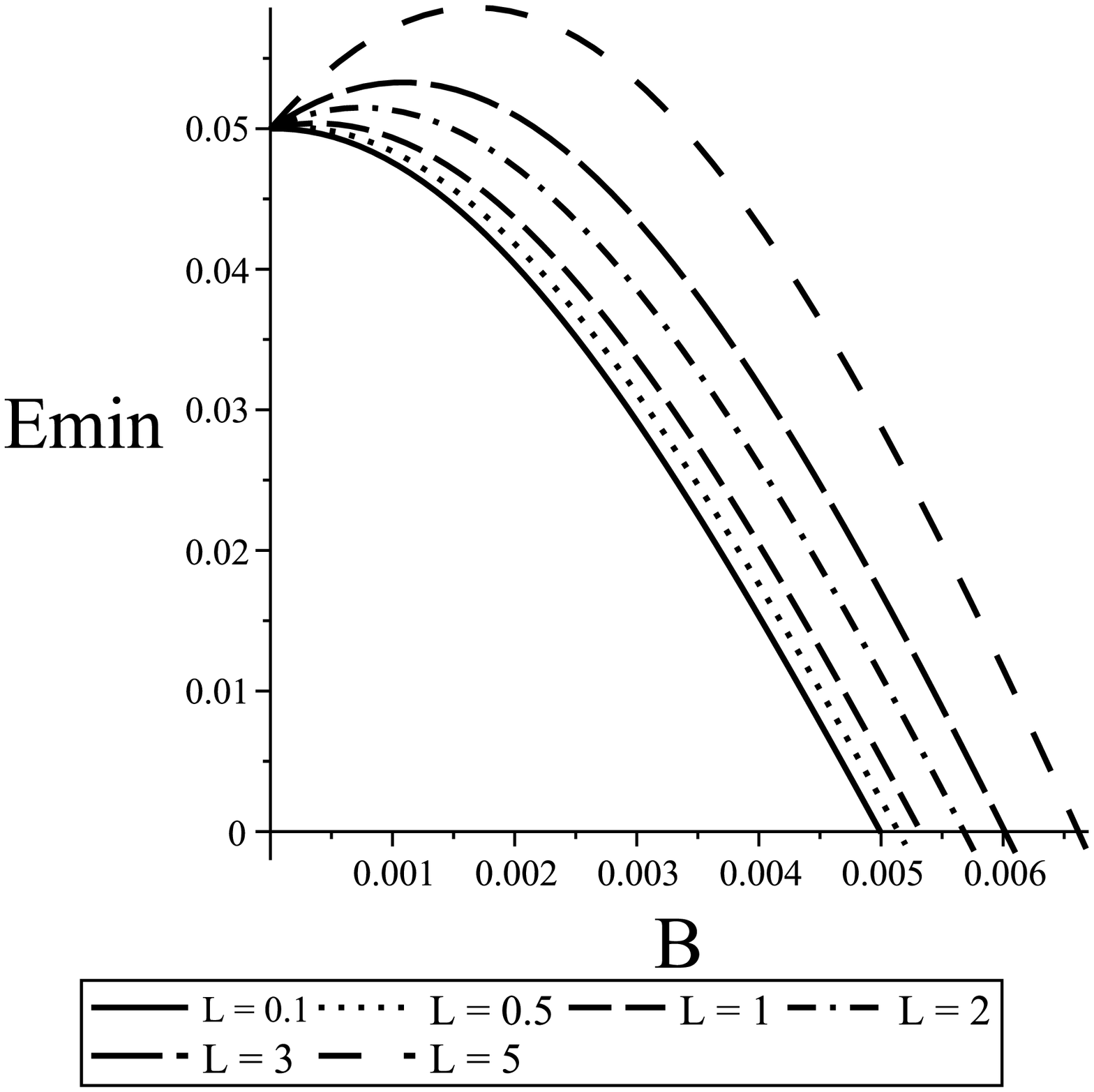}
	\end{center}
	\caption{Plots of $E_{\min}$ for $M=100$, $q=0.05$, and several numerical values of charge parameter $L$.} \label{Fig.EminMRNvarLminL}
\end{figure}

\begin{figure}
	\begin{center}
		\includegraphics[scale=0.4]{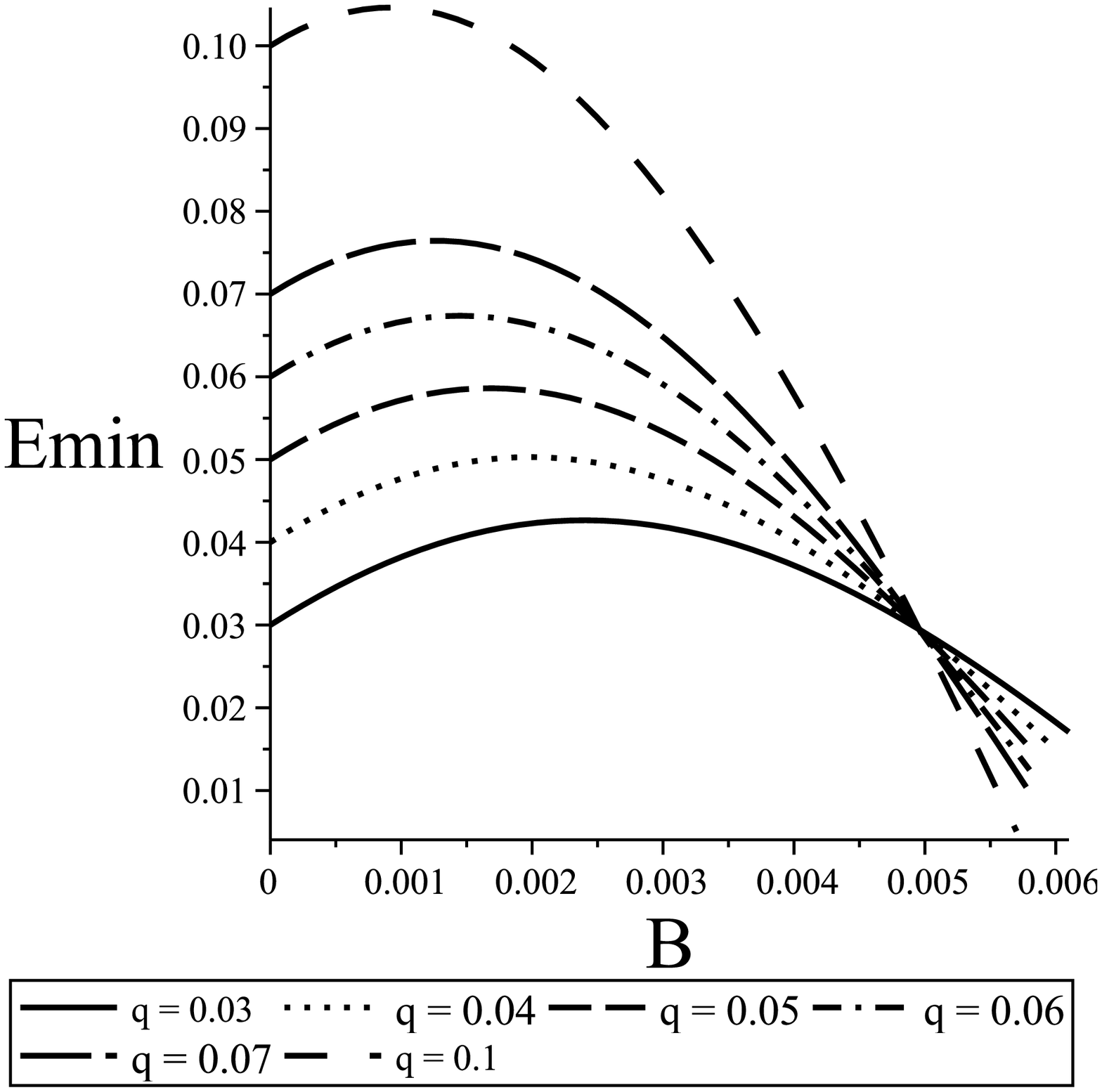}
	\end{center}
	\caption{Plots of $E_{\min}$ for $M=100$, $L=5$, and several numerical values of charge parameter $q$.} \label{Fig.EminMRNvarQminL}
\end{figure}

To support the claim above, we provide numerical plots in Figs. \ref{DelE6plotsLvar} and \ref{DelE6plotsQvar} where the exact energy gap $\Delta E$ in (\ref{EminMRN}) is evaluated for some particular numerical values. In Fig. \ref{DelE6plotsLvar}, we vary the numerical values of test particle angular momentum $L$, while the variation of numerical values for $q$ is done in Fig. \ref{DelE6plotsQvar}. Both plots agree with the claim that a test particle may destroy an extremal magnetized \RN black holes, in a range of external magnetic field strength, depending on the physical properties of black holes and test particles. It is also important to make sure that the particle has non-negative minimum energy $E_{\min}$, which also can be done numerically as in Figs. \ref{Fig.EminMRNvarLminL} and \ref{Fig.EminMRNvarQminL}. Note that the range of $B$ where $E_{\min}$ remains positive is considerably small compared to the range of $B$ for a positive $\Delta E$. This fact should be one of the considerations in determining the values of test particle parameters in the next section where some numerical examples are presented.

Now let us turn our discussion to ask whether the test particle that has chance to destroy the black hole can really reach the horizon, if the particle is released far away from the black hole. This question can be answered by observing the effective potential $V_{eff}$ describing the test particle in the background of magnetized \RN black holes. If $V_{eff} < 0$ everywhere outside the horizon, then one can conclude that the particle described by this $V_{eff}$ can arrive at the horizon from its initial far position. In order to keep the generalities, quantities under discussion except black hole mass can be expressed in term of black hole mass $M$. Following \cite{Wald:1984rg.book}, this effective potential is given by
\be \label{Veff}
V_{eff}  =  - \frac{{\dot r^2 }}{2}\,,
\ee 
where ${\dot r}^2$ is obtained from (\ref{EtoL}), i.e. by solving 
\be \label{rdot.eq}
m^2 g_{\phi \phi } \left( {1 + g_{rr} \dot r^2 } \right) = \left( {\frac{{g_{\phi \phi }^2 }}{{g_{t\phi }^2  - g_{\phi \phi } g_{tt} }}} \right)\left( {E - \frac{{g_{t\phi } }}{{g_{\phi \phi } }}\left( {qA_\phi   - L} \right) + qA_t } \right)^2  - \left( {L - qA_\phi  } \right)^2 \,.
\ee
If we can show that
\be \label{VeffCondition}
V_{eff} < 0\,\,\, \forall ~r > r_+\,,
\ee 
then we can conclude that the test particle can really reach the horizon from far away. We restrict our discussion to the geodesic in equatorial plane only, where in App. \ref{app.equator} we show that such geodesic can occur outside a general magnetized black hole. Accordingly, one can show
\be \label{VeffMBH}
V_{eff}  = \frac{{{\tilde g}_{tt} \left( {m^2 {\tilde g}_{\phi \phi }  + \left( {q{\tilde A}_\phi   - L} \right)^2 } \right) - 2\left( {E + q{\tilde A}_t } \right)\left( {q{\tilde A}_\phi   - L} \right){\tilde g}_{t\phi }  + {\tilde g}_{\phi \phi } \left( {m^2 {\tilde g}_{tt}  + \left( {E + q{\tilde A}_t } \right)^2 } \right)}}{{2m^2 {\tilde g}_{rr} \left( {{\tilde g}_{\phi \phi } {\tilde g}_{tt}  - {\tilde g}_{t\phi }^2 } \right)}}
\ee
where ${\tilde g}_{\mu\nu}$ and ${\tilde A}_\mu$ are the metric tensor and vector fields of the magnetized \RN solution with the conditions $Q=M$ and $\theta = \tfrac{\pi}{2}$. The exact expression of this $V_{eff}$ is considerably complicated, hence for simplicity we prefer not to show it explicitly. Nonetheless, one can verify this effective potential asymptotically, 
\be \label{VeffMRNrtoinf}
\mathop {\lim }\limits_{r \to \infty } V_{eff}  = 0\,,
\ee
and
\be \label{VeffMRNrtozero}
\mathop {\lim }\limits_{r \to 0} V_{eff}  = \infty \,.
\ee

Furthermore, this complexity of $V_{eff}$ (\ref{VeffMBH}) hinders us to perform further analytical investigation to verify the fate of test particle which is approaching black hole from somewhere far away. To guarantee that the black hole can really be reached by the test particle, we need to check whether the effective potential (\ref{VeffMBH}) satisfies the condition (\ref{VeffCondition}). Nonetheless, some reasonable approximations on the magnetic field strength and the test particle properties such as $m$, $q$, and $E$ can be performed which finally yield the effective potential to depend on $M$ and $r$ only. Thence, without loss of generality, we may judge the behavior of effective potential from the leading terms in $V_{eff}$ after approximations. We understand that the test particle properties, in the unit being used, are very small compared to black hole's mass $M$, i.e. $E \ll M$, $m \ll M$, and $q \ll M$. Hence we can have $E \sim \lambda M$, $m \sim \lambda M$, $q \sim \lambda M$, and $L \sim \lambda M^2$ for $\lambda \ll 1$. Here the notation ``$\sim$'' stands for ``in the order of''. For the magnetic fields, even though Figs. \ref{DelE6plotsLvar} and \ref{DelE6plotsQvar} tell us that $\Delta E >0$ for a range of weak $B$ only, here we will consider both the strong and weak magnetic field cases represented by $BM \sim 1$ and $BM \sim \lambda$ respectively.

\begin{figure}
	\begin{center}
		\includegraphics[scale=0.45]{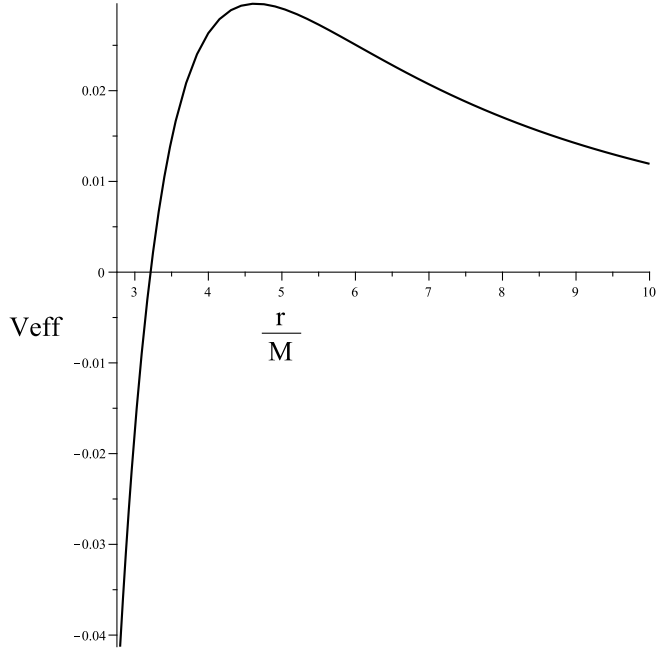}
	\end{center}
	\caption{$V_{eff}$ for a test particle outside a magnetized \RN black hole immersed in strong magnetic fields. The plot is taken for $r\ge 2.8 M$.} \label{Fig.VeffMRNstrong}
\end{figure}

Now let us see the behavior of $V_{eff}$ in these both cases. In the case of weak external magnetic fields, denoted by $B \sim \lambda M^{-1}$, one can show 
\be\label{VeffweakMRN}
V_{eff,weak}  = \frac{{\left( {r - M} \right)^2 M^2 }}{{2r^4 }}
\ee 
after the limit $\lambda \to 0$ is taken. In this weak field case, it is easy to see that particle cannot reach the horizon released at some points far away from black hole since $V_{eff,weak} > 0$ outside the horizon. In case of strong external magnetic field, the effective potential takes the form
\[
V_{eff,strong}  = \frac{8}{{1681r^4 {\cal F}\left(r\right)}}\left\{ {1600{M}^{10}-3200{M}^{9}r+16480{r}^{2}{M}^{8}+18288{M}^{7}{r}^{3}-81808{M}^{6}{r}^{4}} \right.
\]
\be\label{VeffMRNstrong} 
\left.~~~~~~~~~ {-107924{r}^{5}{M}^{5}+9156{r}^{6}{M}^{4}-7826{r}^{7}{M}^{3}+4237{r}^{8}{M}^{2}-648{r}^{9}M+324{r}^{10}} \right\}
\ee
where ${\cal F} \left(r\right) = {256{M}^{8}+256{M}^{6}{r}^{2}+96{M}^{4}{r}^{4}+16{M}^{2}{r}^{6}+{r}^{8}}$. Obviously, it is not easy to tell whether the test particle can reach the horizon in strong external magnetic case based on the approximate effective potential above. Alternatively, the numerical plot in Fig. \ref{Fig.VeffMRNstrong} illustrating the effective potential (\ref{VeffMRNstrong}) helps us to conclude that, even in a strong magnetic field background, the test particle can never reach the horizon.

\subsection{Test particles in the magnetized Kerr spacetime}\label{ss.TK}

Studying the behavior of a test particle in the background of extremal magnetized Kerr black holes can be done by using the similar method performed previously. From (\ref{EtoL}), the corresponding minimum energy required by a test particle in reaching the extremal magnetized Kerr black hole's horizon is given by
\be\label{EminKerr} E_{\min } = \frac{{L\left( {1 - B^4 M^4 } \right)}+{4qM^4 B^3\left(1+2B^4 M^4\right)}}{{2M\left( {1 + B^4 M^4 } \right)}}\,,
\ee
where we have imposed $r=r_+$. This minimum energy requires $B \le M^{-1}$ to be non-negative\footnote{For the numerical value $M=100$ which is used in this paper, it means $B \le 0.01$.}. Similar to the prescription employed in the previous subsection, the upper bound for test particle's energy can be obtained from (\ref{viol.ext}), where the charge parameter of black hole $Q$ is set to be zero. The resulting equation is  
\be\label{EmaxMK}
E_{\max }  = \frac{1}{2}\sqrt {2\left( {\tilde Q_{K} +q } \right)^2  + 2\sqrt {\left( {\tilde Q_{K} + q } \right)^4  + 4\left( {\tilde J_{K} } +L \right)^2 } }  - \tilde M_{K} \,.
\ee 
where ${\tilde M}_K$, ${\tilde Q}_K$, and ${\tilde J}_K$ are given in (\ref{massMK}), (\ref{QMK}), and (\ref{JMK}) respectively. In terms of the physical parameters in the seed solution, equation (\ref{EmaxMK}) can be read as 
\be 
E_{\max }  = \frac{{\sqrt 2 }}{2}\left( {\sqrt {\left( {2M^2 B + q} \right)^4  + 4\left( {M^2  + L - M^6 B^4 } \right)^2 }  + \left( {2M^2 B + q} \right)^2 } \right)^{\frac{1}{2}}  - M - M^3 B^2 \,,
\ee
where the extremal condition $J=M^2$ has been considered.

\begin{figure}
	\begin{center}
		\includegraphics[scale=0.4]{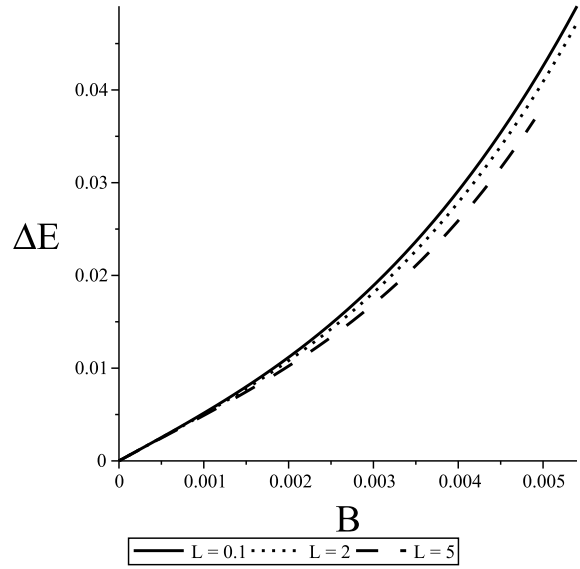}
	\end{center}
	\caption{Plots of $\Delta E = E_{\max} -E_{\min}$ for $M=100$, $q=0.05$, and several numerical values of charge parameter $L$.} \label{Fig.DelEMKvarL}
\end{figure}

\begin{figure}
	\begin{center}
		\includegraphics[scale=0.4]{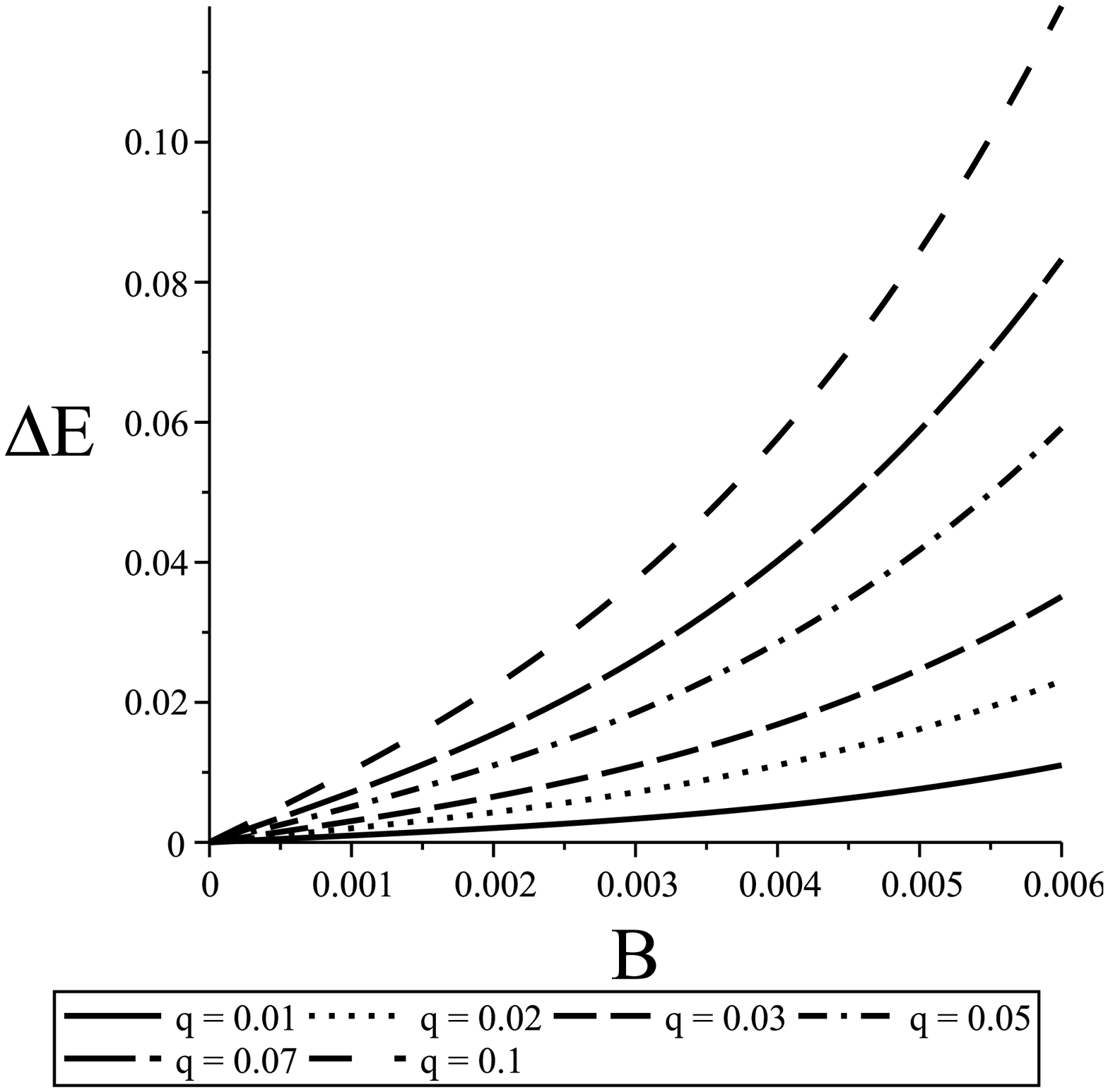}
	\end{center}
	\caption{Plots of $\Delta E = E_{\max} -E_{\min}$ for $M=100$, $L=1$, and several numerical values of charge parameter $q$.} \label{Fig.DelEMKvarQ}
\end{figure}

As in the previous subsection, if $\Delta E \equiv E_{\max}-E_{\min} > 0$ then the particle has a possibility to break the horizon. However, evaluating $\Delta E$ exactly would be tedious, thence we could Taylor expand $\Delta {\cal E} \equiv {\cal E}_{\max} - E_{\min}$ for small $B$, where
\be 
{\cal E}_{\max} = \frac{1}{2}\sqrt {2\left( {\tilde Q_{K}} \right)^2  + 2\sqrt {\left( {\tilde Q_{K}} \right)^4  + 4\left( {\tilde J_{K} } \right)^2 } }  - \tilde M_{K}  < E_{\max}\,.
\ee 
The Taylor expansion for $\Delta {\cal E}$ can be found to be
\be \label{DelEappMK}
{\Delta {\cal E}} \approx  - \frac{L}{{2M}} + 2M^3 qB^3 \,.
\ee 
This result hints the possibility of positive energy gap for the test particle, which then signs the potential of black hole horizon to be broken if such test particle is captured. To support this conclusion, which should be valid in the weak $B$ regime only, we provide the numerical plots given in Figs. \ref{Fig.DelEMKvarL} and \ref{Fig.DelEMKvarQ} where several different possible values for test particle properties are considered. We can see from these plots, for each variations of $L$ and $q$ considered, that the energy gap $\Delta E$ grows nonlinearly as $B$ increases. This nonlinearity might be the one that is predicted by $\Delta {\cal E} \sim B^3$ in (\ref{DelEappMK}).

\begin{figure}
	\begin{center}
		\includegraphics[scale=0.4]{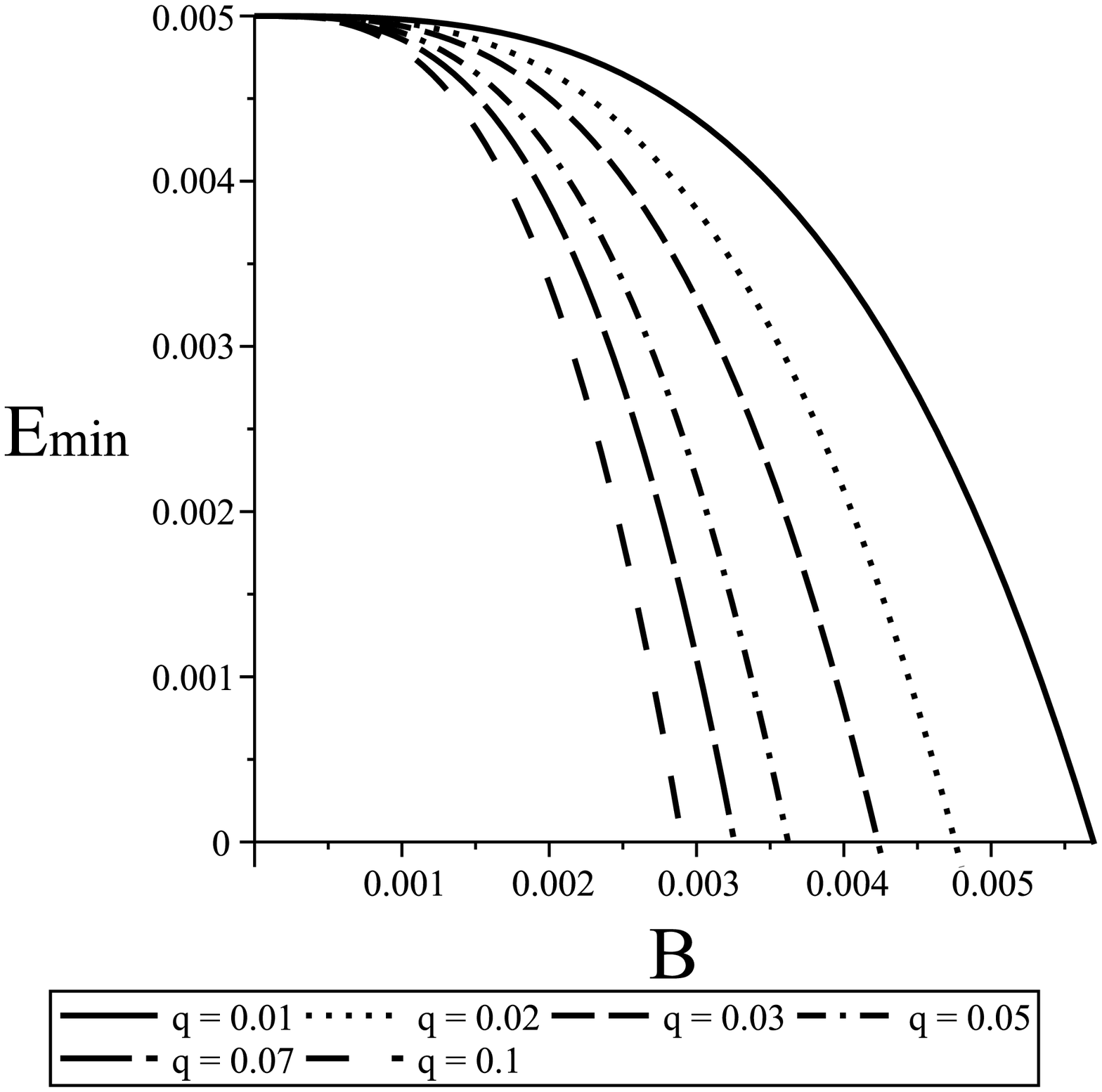}
	\end{center}
	\caption{Plots of $E_{\min}$ for $M=100$, $L=1$, and several numerical values of charge parameter $q$.} \label{Fig.EminMKvarQ}
\end{figure}

\begin{figure}
	\begin{center}
		\includegraphics[scale=0.4]{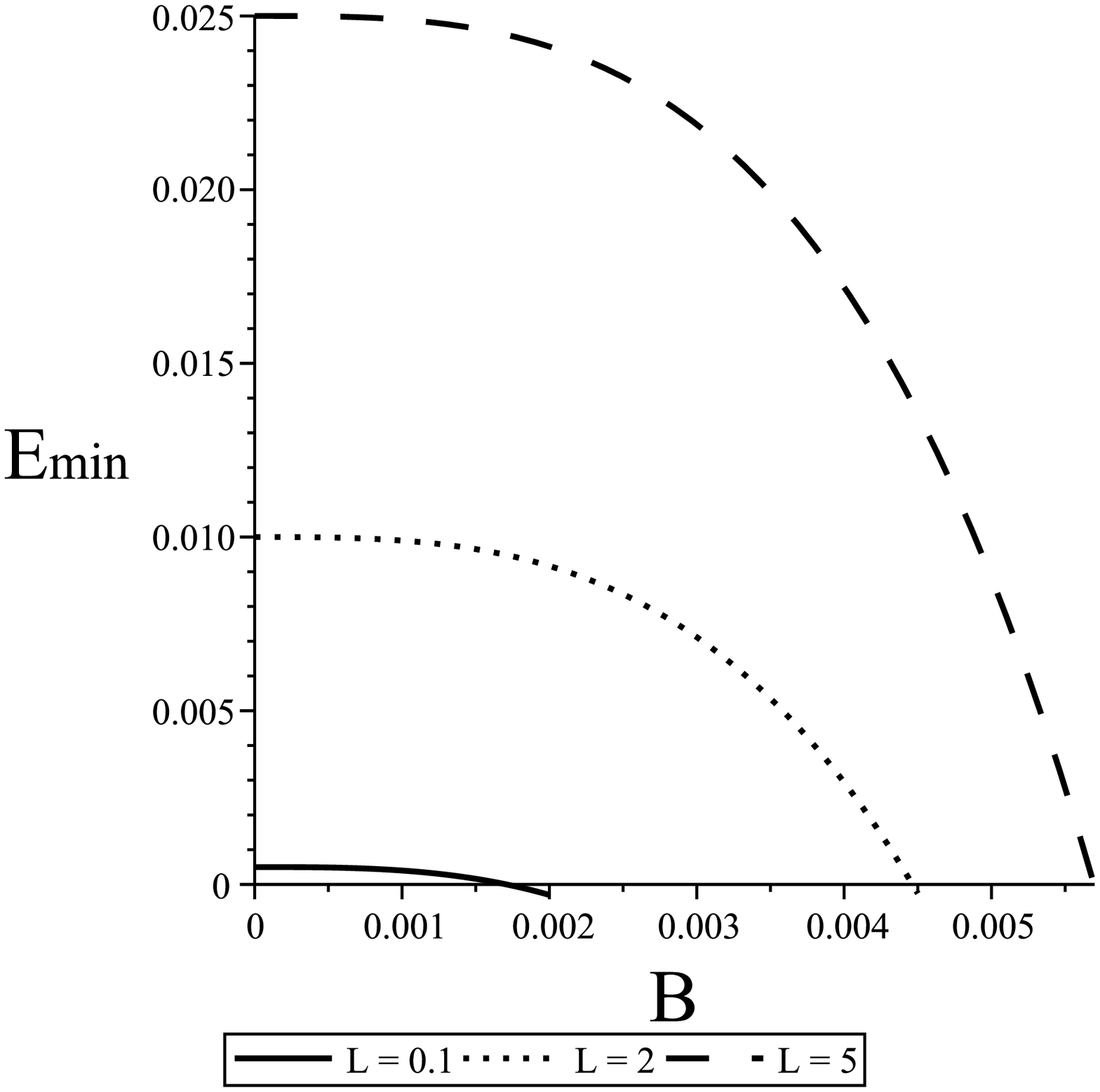}
	\end{center}
	\caption{Plots of $E_{\min}$ for $M=100$, $q=0.05$, and several numerical values of charge parameter $L$.} \label{Fig.EminMKvarL}
\end{figure}

However, there is a constraint for the magnetic field strength $B$, that the minimum energy required by the test particle must be non-negative. Thence, we cannot consider all range of $B$ in our thought experiment. Figs. \ref{Fig.EminMKvarL} and \ref{Fig.EminMKvarQ} provide us the corresponding information of the ``allowed'' $B$ for each cases considered in Figs. \ref{Fig.DelEMKvarL} and \ref{Fig.DelEMKvarQ}. From these plots, one can learn that considering $B = 0.01$ as the upper bound of magnetic field suit the constraint for positive $E_{\min}$. Note that for the rest of this paper, we use $L =5$ and $q=0.1$ as the maximum values of the test particle's angular momentum and charge.

Having observed the possibility of a test particle to destroy an extremal magnetized Kerr black hole from the energy gap point of view, now let us check whether such particle released from far away can really arrive at the horizon. Prescription in doing this is exactly the same as that in the previous subsection, i.e. by looking at the corresponding effective potential $V_{eff}$. The expression of $V_{eff}$ can be obtained from eq. (\ref{VeffMBH}) after substituting ${\tilde g}_{\mu\nu}$ and ${\tilde A}_{\mu}$ that belong to the extremal magnetized Kerr solutions. Again, the full expression of the $V_{eff}$ in this case is quite complicated, therefore it would be difficult to draw any conclusion directly. Nevertheless, the same approximation method as we performed in discussing magnetized \RN can also be employed in this case. We can set $m\sim\lambda M$, $q\sim\lambda M$, and $L\sim\lambda M^2$. We also consider the cases of weak and strong magnetic fields, even though Figs. \ref{Fig.EminMKvarL}
and \ref{Fig.EminMKvarQ} suggest that $E_{\min}$ may be negative if $B$ is too strong. As in the previous case, $B\sim \lambda M^{-1}$ and $B\sim M^{-1}$ represent the weak and strong magnetic field cases respectively.

\begin{figure}
	\begin{center}
		\includegraphics[scale=0.4]{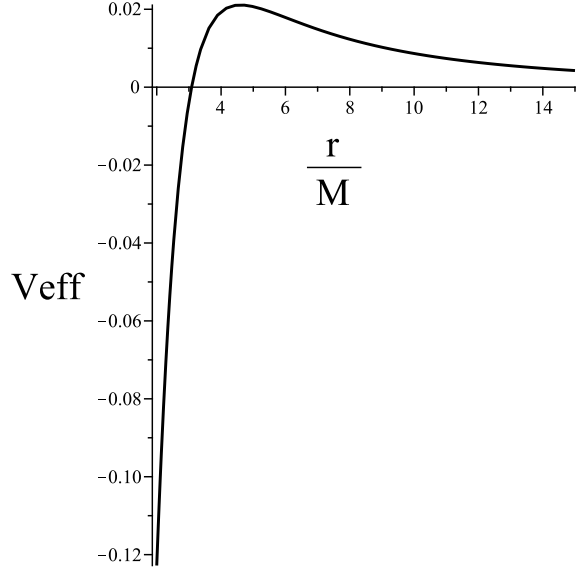}
	\end{center}
	\caption{$V_{eff,strong}$ for $r \ge 2.5 M$.} \label{Fig.VeffMKstrong}
\end{figure}

After taking the limit $\lambda \to 0$, we obtain
\be \label{VeffMKweak}
V_{eff,weak} = \frac{1}{2}{\frac {M \left( M-2r \right) }{{r}^{2}}}\,,
\ee 
and
\[ 
V_{eff,strong} = \frac{{M}^{2}}{8 {\cal F}_2 \left(r\right)^2} \left\{ 1432{M}^{10}-2298{M}^{9}r-7223{M}^{8}{r}^{2}-3024{M}^{7}{r}^{3}\right.
\]
\be \label{VeffMKstrong}\left. ~~~~
-2612{M}^{6}{r}^{4}-3012{M}^{5}{r}^{5}+438{M}^{4}{r}^{6
}-224{r}^{7}{M}^{3}+148{M}^{2}{r}^{8}-18M{r}^{9}+9{r}^{10}
\right\}
\ee  
where ${\cal F}_2 \left(r\right) = 2{M}^{3}+5{M}^{2}r+{r}^{3}$. Nevertheless, telling whether the test particle can reach the black hole by reading $V_{eff,strong}$ in (\ref{VeffMKstrong}) is not so easy. However, Fig. \ref{Fig.VeffMKstrong} can help us to do the job, where we learn that in strong magnetic field the test particle cannot arrive at the horizon from far away.

Interestingly, eq. (\ref{VeffMKweak}) hints the possibility of a test particle to destroy a weakly magnetized black hole. This could be a fascinating result, since the black hole immersed in weak magnetic fields may have some relevance with the astrophysical black holes in the sky. However, one can perform a further check on this result by employing a Taylor expansion for small $B$. Definitely this is allowed since the case we are considering here is the weak field condition. By setting $m\sim E\sim q\sim \lambda M$, and $L\sim \lambda M^2$, but let $B$ parameter unchanged in the full expression of $V_{eff}$, and after taking the limit $\lambda \to 0$ one can get an effective potential ${\tilde V}_{eff} \left(r,M,B\right)$. Taylor expanding this ${\tilde V}_{eff} \left(r,M,B\right)$ up to the second order in $B$ yields
\be 
\tilde V_{eff} \left( {r,M,B} \right) \approx  - \frac{{\left( {2r - M} \right)M}}{{2r^2 }} + \frac{1}{2}\left( {\frac{{r - M}}{r}} \right)^2 B + \left( {\frac{3}{4}+{\frac {3{r}^{2}}{8{M}^{2}}}+{\frac {r}{4M}}+{\frac {7{M}^{2}}{8{r}^{2}}}-{\frac {{M}^{3}}{4{r}^{3}}}
} \right)B^2 \,,
\ee 
which can be positive as $r$ increases\footnote{This will be verified in the next section where some numerical evaluations of the exact $V_{eff} $ are given for some particular numerical vaues of $L$, $q$, $m$, $E$ and $B$.}. Accordingly, we can claim that the test particle still cannot reach the horizon in the case of weak magnetic fields.

\section{Numerical examples}\label{s.3}

\begin{figure}
	\begin{center}
		\includegraphics[scale=0.4]{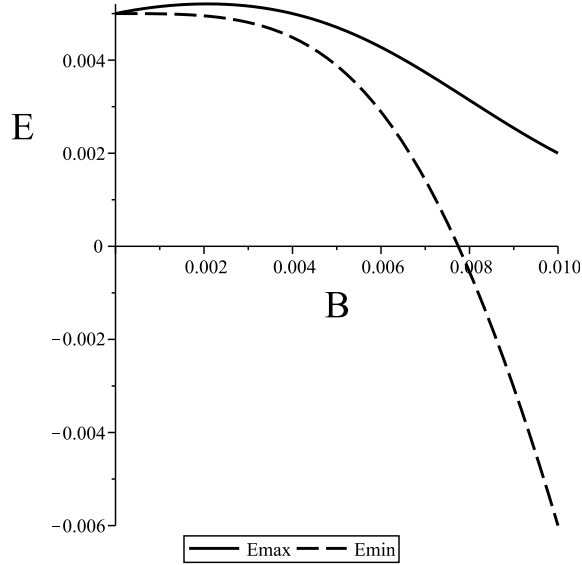}
	\end{center}
	\caption{Example of $E_{\max}$ and $E_{\min}$ for a test particle in the background of magnetized Kerr spacetime as functions of $B$.} \label{Fig.NumEmaxEminKerr}
\end{figure}

We have seen in the previous section, after employing some approximations, the test particle which may break the horizon of magnetized black holes cannot really arrive at that surface. This is because, even though the energies of test particles are in between $E_{\max}$ and $E_{\min}$, the corresponding effective potentials sign the impossibility of this particle to arrive at the horizon. In this section we provide some numerical examples of the possible effective potentials which are related to the test particle in the background of magnetized \RN and Kerr black holes. 

\begin{figure}
	\begin{center}
		\includegraphics[scale=0.6]{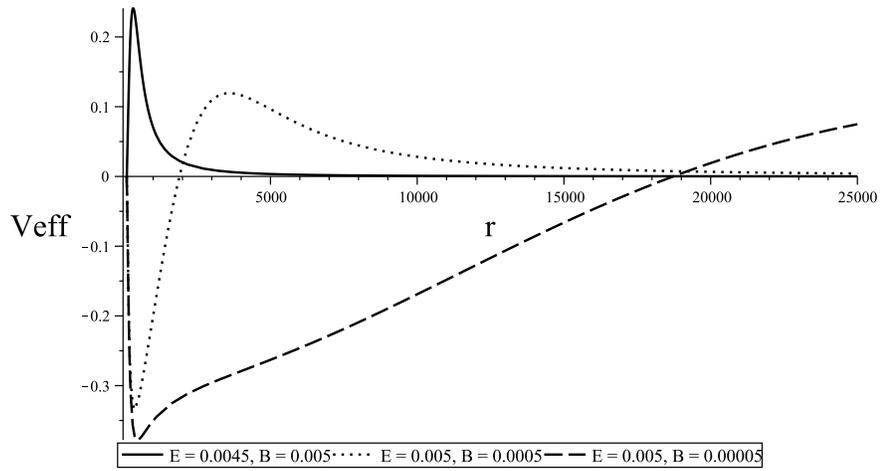}
	\end{center}
	\caption{Examples of $V_{eff}$ for several cases of test particles in magnetized Kerr background with distinguished physical properties.} \label{Fig.NumVeffKerrWeak2}
\end{figure}

By taking $q=0.002$, $m=0.004$, $L=1$, and $M=100$, the maximum and minimum energies are given in Fig. \ref{Fig.NumEmaxEminKerr}. By increasing the magnitudes of $q$ and $M$, the maximum energy can be less than the particle mass, which is irrelevant to be considered. Moreover, from Fig. \ref{Fig.NumEmaxEminKerr}, the minimum energy is positive for $B \lesssim 0.0075$ which is near to the regime of strong magnetic fields. Therefore, in Fig. \ref{Fig.NumVeffKerrWeak2}, we consider the cases of $B=5\times 10^{-3}$, $B=5\times 10^{-4}$, and $B=5\times 10^{-5}$, for the test particle considered in Fig. \ref{Fig.NumEmaxEminKerr}. If one consider the smaller $B$, since it looks Fig. \ref{Fig.NumVeffKerrWeak2} indicates the possibility of $V_{eff} <0~\forall~r>M$ in such consideration, eventually the curve $V_{eff}$ will have positive value in the large radii. Definitely, we have to make sure that $\Delta E >0$ and $E_{min} >0$ in changing the parameters we are using. These examples are in agreement to the conclusions drawn in the previous section, where even though the test particle may fulfill the gap energy condition, it can never reach the horizon.

\begin{figure}
	\begin{center}
		\includegraphics[scale=0.6]{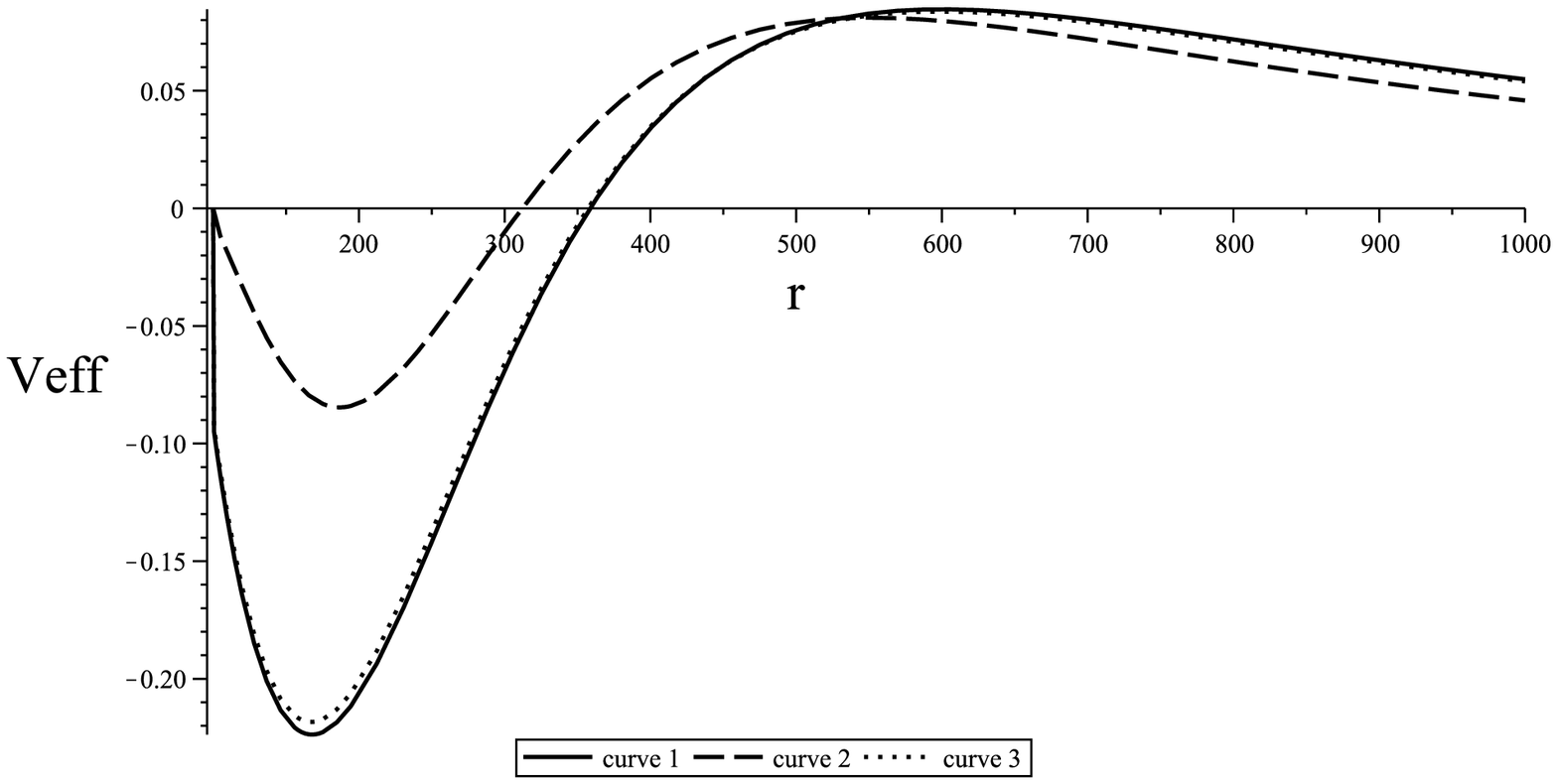}
	\end{center}
	\caption{Examples of $V_{eff}$ for several cases of test particles in magnetized \RN background with distinguished physical properties given in Table \ref{Tab.testMRN}.} \label{Fig.NumVeffRNWeak}
\end{figure}

Now, let us turn to the discussion of magnetized \RN. Compared to the magnetized Kerr case, there is a narrow range of variations that can be made to provide the numerical examples, i.e. test particles with particular properties in the magnetized \RN spacetime with a specific magnetic field strength $B$, which obey the energy gap condition $\Delta E > 0$. One should also keep in mind that $q < m < E$. However, some examples that fulfill these required conditions are given in Table \ref{Tab.testMRN}, and the corresponding $V_{eff}$ plots are given in Fig. \ref{Fig.NumVeffRNWeak}. The numerical results presented in Fig. \ref{Fig.NumVeffRNWeak} support the claim given in the previous section, where destroying an extremal magnetized \RN black hole is impossible. A test particle may have an energy that leads to violation of the extremal bound, but it can never fall into the black hole.

\begin{center}
	\begin{table}
		\begin{tabular*}{0.75\textwidth}{@{\extracolsep{\fill} } |c | c | c | r | c| }
			\hline
			curve 1 & q = 0.05 & m = 0.052 & E = 0.054 & L = 5 \\
			\hline 
			curve 2  & q = 0.02 &  m = 0.03 & E = 0.0325 & L = 5\\
			\hline
			curve 3  & q = 0.01 & m = 0.0105 & E = 0.0108 & L = 1 \\
			\hline
		\end{tabular*}\caption{Physical properties of several test particles.}\label{Tab.testMRN} 
	\end{table}
\end{center}

\section{Discussion and conclusions}\label{s.discussion}

Now let us conclude the works presented in this paper. We have shown that, even though the presence of magnetic field yields a test particle to have energies allowing the magnetized Kerr and \RN black holes destruction, the corresponding effective potentials indicate that this particle can never reach the horizon. This conclusion is drawn using two approaches. Firstly, we employ the semi-analytic study on the associated effective potentials for each cases presented in section \ref{s.2}, and secondly by providing some numerical plots in section \ref{s.3}. Results obtained in both methods are in agreement, and tell us that the extremal magnetized \RN and Kerr black holes cannot be destroyed by throwing a test particle in equatorial plane according to Wald's gedanken experiment \cite{waldAP1974}. Our analysis is in the same fashion as the studies presented in \cite{Gao:2012ca,Siahaan:2015ljs}, where the self-force, self-energy, and radiative effects are neglected. Nevertheless, we are sure that taking these effects into account would not change the present conclusion, on the possibility of destroying the magnetized black holes. This is because, on the contrary, these effects are normally used to restore the cosmic censorship rather than to show the fragility of a black hole against a test particle.

As it is pointed out in Wald's book \cite{Wald:1984rg.book}, that studying the solutions in Einstein-Maxwell theory can be of great importance, pursuing further the works presented in this paper might be interesting. For example, the discussions performed in this paper are confined to the case of extremal black holes only. Nonetheless, Hubeny \cite{Hubeny:1998ga}, Sotiriou and Jacobson \cite{Jacobson:2009kt}, and many others \cite{Gao:2012ca,Siahaan:2015ljs,SaaPRD} have shown that the extremality can be jumped if the initial condition of the black holes is near-extremal. Then it is straightforward question to ask what happens to a near-extremal magnetized black hole perturbed by a test particle. Can this test particle have $\Delta E >0$ and experience $V_{eff} <0\,\,\forall\, r > r_+$ ? It might sound trivial, but considering that the mathematical expressions of fields in the magnetized black holes are quite involved, pursuing this work could be appealing. In addition to that, extending this work to the case of extremal magnetized Kerr-Newman can also be interesting.

\section*{Acknowledgement}

This work is supported by LPPM of Parahyangan Catholic University. I thank Dr. Paulus C. Tjiang of UNPAR Physics Dept. for the discussions and encouragement. I also thank the anonymous referee for his/her comments.

\appendix

\section{Equatorial plane in the magnetized spacetimes}\label{app.equator}

The test particle's dynamics in the magnetized spacetimes discussed above obeys the Euler-Lagrange equation with the Lagrangian
\be 
{\cal L} = \frac{m}{2}g_{\alpha \beta } \dot x^\alpha  \dot x^\beta   + qA_\alpha  \dot x^\alpha  \,.
\ee 
To show that the case where our test particle moving only in the equatorial plane, namely at $x=0$, we have to show that the Euler-Lagrange which corresponds to this case is satisfied. One can show that for the magnetized Kerr and \RN backgrounds,
\be 
\left. {\frac{{\partial {\cal L}}}{{\partial x}}} \right|_{x = 0}  = \frac{m}{2}\dot x^\alpha  \dot x^\beta  \left( {\left. {\frac{{\partial g_{\alpha \beta } }}{{\partial x}}} \right|_{x = 0} } \right) + q\dot x^\alpha  \left( {\left. {\frac{{\partial A_\alpha  }}{{\partial x}}} \right|_{x = 0} } \right) = 0\,,
\ee
since 
\be 
\left. {\frac{{\partial g_{\alpha \beta } }}{{\partial x}}} \right|_{x = 0} = 0
\ee
and
\be 
\left. {\frac{{\partial A_\alpha  }}{{\partial x}}} \right|_{x = 0} = 0
\ee
for these two cases. Furthermore, it is easy to see that
\be 
\frac{{\partial {\cal L}}}{{\partial \dot x}} = mg_{xx} \dot x
\ee
which vanishes for a fixed $x$. Based on these facts, now we see that the case of a test particle moving only in the equatorial plane satisfies the Euler-Lagrange equation for that particle.

\section{Mass and energy in magnetized spacetime}\label{app.Mass}

The fact that magnetized spacetime by Ernst and Wild are not asymptotically flat nor (A)dS requires a special treatment to define mass and energy in this background. Several attempts have been reported in literature, for example \cite{Booth:2015nwa,Gibbons:2013dna,Astorino:2016hls}, where the authors managed to get mass definition of an object immersed in external magnetic fields. In particular, the magnetized black hole mass reported in \cite{Astorino:2016hls} is obtained by employing the Barnich-Brandt method \cite{Barnich:2001jy,Barnich:2003xg,Barnich:2007bf} and integrability condition. The result for black hole mass in \cite{Astorino:2016hls} agrees to that in \cite{Booth:2015nwa} where the authors use the isolated horizon technique. This appendix is devoted to highlight the results reported in \cite{Astorino:2016hls}, which are used in this paper to define some physical quantities of a magnetized black hole. 

Using the Barnich-Brandt method and integrability, the mass of magnetized Kerr-Newman black holes can be shown as \cite{Astorino:2016hls}
\[
\tilde M  = \left[ M^2  + 2JQB + \left( {2J^2  + \frac{3}{2}M^2 Q^2  - Q^4 } \right)B^2  + JQ\left( {2M^2  - \frac{3}{2}Q^2 } \right)B^3    \right.
\]
\be\label{massMKN}
\left. + \left( {J^2 M^2  - \frac{1}{2}J^2 Q^2  + \frac{1}{{16}}M^2 Q^4 } \right)B^4 \right]^{\tfrac{1}{2}}\,.
\ee 
In the expression above, $M$, $J$, and $Q$ are the mass, angular momentum, and charge parameters of black holes respectively, i.e. the parameters which are brought from the seed Kerr-Newman solution. Moreover, it is showed that the law of black hole mechanics for a magnetized Kerr-Newman black hole is the same to that of Kerr-Newman one in standard form, i.e.
\be \label{dMmKN}
\delta \tilde M = T_H\delta S + \Omega \delta \tilde J + \Phi \delta \tilde Q\,,
\ee 
where $\tilde J$ and $\tilde Q$ are the angular momentum and charged of magnetized Kerr-Newman black hole\footnote{The Hawking temperature $T_H$ is not necessary to be expressed with ``tilde'', since the this temperature does not change when the black hole is immersed in external magnetic field \cite{Booth:2015nwa}.}, namely \cite{Astorino:2016hls}
\be\label{JMKN}
\tilde J = J - Q^3 B - \frac{3}{2}JQ^2 B^2  - \frac{Q}{4}\left( {8J^2  + Q^4 } \right)B^3  - \frac{J}{{16}}\left( {16J^2  + 3Q^4 } \right)B^4 \,,
\ee
and
\be \label{QMKN}
\tilde Q = Q + 2JB - \frac{{Q^3 B^2 }}{4}\,,
\ee 
respectively. In particular, the corresponding conserved quantities in a magnetized \RN spacetime can be obtained from (\ref{massMKN}), (\ref{JMKN}), and (\ref{QMKN}) by setting $J=0$ in each of these expressions. Similarly, conserved quantities in a magnetized Kerr spacetime are achieved by setting $Q=0$ in the three $\tilde M$, $\tilde J$, and $\tilde Q$ for magnetized Kerr-Newman above.

Dealing with the concept of energy in magnetized spacetime, whose asymptotics are not flat nor (A)dS, is more delicate. In an asymptotically flat spacetime, the standard textbook method\footnote{See e.g. \cite{Wald:1984rg.book}.} in defining energy can be done by using the timelike Killing vector which is normalized at infinity. However, in the case of magnetized spacetime, this method cannot be employed. Nevertheless, since we are interested in the study of the change of black hole mass after capturing a test particle, we can relate the energy of test particle with $\delta{\tilde M}$ which follows (\ref{dMmKN}). So, the particle's energy $E$ discussed in section \ref{s.2} can be identified as $\delta {\tilde M}$, which yields the post-infalling mass of the magnetized black holes is ${\tilde M}+\delta {\tilde M}$.

Furthermore, to guarantee the entropy for magnetized Kerr-Newman black holes to be real, the following relation must be fulfilled \cite{Astorino:2016hls}
\be \label{extremeMKN}
0 \le \tilde M^4  - \tilde Q^2 \tilde M^2  - \tilde J^2  = M^2 \left( {M^2  - \frac{{J^2 }}{{M^2 }} - Q^2 } \right)\left( {\left| {\Lambda _{RN,0} } \right|^2  + \left| {\Lambda _{K,0} } \right|^2  + 2JQB^3 } \right) \,,
\ee 
where ${\Lambda _{RN,0} }$ and ${\Lambda _{K,0} }$ are given in (\ref{Lambda0RN}) and (\ref{Lambda0K}) respectively. Eventually, eq. (\ref{extremeMKN}) is just the non-existence of naked singularity in Kerr-Newman spacetime. Therefore, since the conserved quantities in the magnetized spacetime under consideration in this paper is $\tilde M$, $\tilde Q$, and $\tilde J$, an inequality which represents the violation of eq. (\ref{extremeMKN}) after capturing the test particle reads
\be \label{viol.ext}
\left( {\tilde M + E} \right)^2  < \frac{{\left( {\tilde Q + q} \right)^2  + \sqrt {\left( {\tilde Q + q} \right)^2  + 4\left( {\tilde J + L} \right)^2 } }}{2}\,,
\ee
where $q$, $L$ and $E$ are the test particle's electric charge, angular momentum, and energy respectively. In writing (\ref{viol.ext}), we have identified the change of black hole mass $\delta {\tilde M}$ which obeys (\ref{dMmKN}) to be the test particle's energy. By using the appropriate inequalities coming from (\ref{viol.ext}), one can obtain the maximum energy of the test particle which leads to the black hole's destruction in subsections \ref{ss.TRN} and \ref{ss.TK}.

\end{document}